\documentclass[reprint,a4paper,superscriptaddress,amsmath,amssymb,aps,pra,11pt,floatfix,tightenlines]{revtex4-2}

\usepackage{graphicx}% Include figure files
\usepackage{xcolor}
\usepackage{dcolumn}% Align table columns on decimal point
\usepackage{bm}% bold math
\usepackage{floatrow} %setting the caption beside figure
\usepackage[all]{nowidow}
\usepackage{siunitx}[=v2]
\usepackage[hidelinks]{hyperref}
\usepackage{ulem}
\makeatletter
\def\maketitle{
\@author@finish
\title@column\titleblock@produce
\suppressfloats[t]}
\makeatother

\newcommand{\pr}{$\varphi_{\omega-2\omega}$}

\begin{document} 
\title{
Optical control of electrons in a Floquet topological insulator}
\author{Daniel M. B. Lesko}
\thanks{These authors contributed equally to this work.}
\affiliation{Department of Physics, Friedrich-Alexander-Universität Erlangen-Nürnberg (FAU), Erlangen, Germany}
\author{Tobias Weitz}
\thanks{These authors contributed equally to this work.}
\affiliation{Department of Physics, Friedrich-Alexander-Universität Erlangen-Nürnberg (FAU), Erlangen, Germany}
\author{Simon Wittigschlager}
\affiliation{Department of Physics, Friedrich-Alexander-Universität Erlangen-Nürnberg (FAU), Erlangen, Germany}
\author{Weizhe Li}
\affiliation{Department of Physics, Friedrich-Alexander-Universität Erlangen-Nürnberg (FAU), Erlangen, Germany}
\author{Christian Heide}
\affiliation{Department of Physics, Friedrich-Alexander-Universität Erlangen-Nürnberg (FAU), Erlangen, Germany}
\affiliation{Department of Physics, University of Central Florida, Orlando, Florida 32816, USA}
\affiliation{CREOL, The College of Optics and Photonics, University of Central Florida, Orlando, Florida 32816, USA}
\author{Ofer Neufeld}
\affiliation{Schulich Faculty of Chemistry, Technion - Israel Institute of Technology, Haifa, Israel}
\affiliation{Max Planck Institute for the Structure and Dynamics of Matter, Hamburg, Germany}
\author{Peter Hommelhoff}
\affiliation{Department of Physics, Friedrich-Alexander-Universität Erlangen-Nürnberg (FAU), Erlangen, Germany}	
\affiliation{Fakult\"at f\"ur Physik, Ludwig-Maximilians-Universit\"at, M\"unchen, Germany}	
\date{\today}

% Insert the title and author list
\maketitle

\onecolumngrid
\vspace{-1.0cm}
\textbf{
Light-dressed materials hold enormous potential for generating new electronic properties. 
The band structure resulting from light-dressing can exhibit starkly different quantum and topological phenomena.
So far, optical control of charge within a light-dressed band structure has been elusive.
Here, we demonstrate optical control of electrons in light-dressed graphene. By focusing circularly polarized femtosecond laser pulses at 1550\,nm on monolayer graphene, we generate a Floquet topological insulator (FTI). With a phase-locked second harmonic field, we dynamically control electrons in this FTI state. For the first time, we observe photocurrent circular dichroism, the all-optical anomalous Hall effect, and FTI valley-polarized currents. The photocurrents show strong sub-cycle phase-sensitivity, opening the door to ultrafast control within topologically protected electronics (topotronics), spectroscopy, and attosecond physics in novel quantum materials.
}

\twocolumngrid

\noindent

When light interacts with a material, the electronic band structure is modified. 
The oscillatory and periodic nature of the light field generates new light-dressed states, described by the Floquet theorem (the time analogue to the Bloch theorem), generating periodic bands spaced by the dressing energy. 
These new quasi-static light-dressed states can exhibit new quantum and topological properties, transformed from their original material eigenstates by the amplitude, phase, and symmetry of the optical dressing field. A particularly exciting class of these light-dressed materials are Floquet topological insulators, where a topologically trivial material is transformed into a topological insulator by light-dressing. Floquet topological insulators represent out-of-equilibrium systems showing intriguing topological properties not present in the undressed material.  
While first predicted for optically driven solids~\cite{Oka2009,Lindner2011,Oka2019,Rudner2020}, most insights in Floquet topological physics have been obtained with photonic waveguides and cold atoms in optical lattices~\cite{Rechtsman2013,Aidelsburger2013,Jotzu2014,Goldman2014,Goldman2016,Zhu2024}.
Experiments in synthetic systems have shown the fascinating ability to design and control material properties with the Floquet Hamiltonian, elucidating both static and time-dependent phenomena~\cite{Desbuquois2017,Wintersperger2020}.

In solids, Floquet topological insulators (FTIs) promise to offer high tunability with the driving laser field properties and the choice of the underlying material system. 
While theory suggests that Floquet dressing can synthesize almost arbitrary material band structures with varying topological phenomena~\cite{Sato2019,Oka2019,Rudner2020}, Floquet topological states in real materials have yet to be directly observed and controlled.
Pioneering works observed Floquet-Bloch bands on a topological surface state~\cite{Wang2013}, as well as engineered bands in a bulk semiconductor~\cite{Zhou2023}.
In a wide bandgap semiconductor, ultrafast Floquet physics can be used to explain phenomena like the enhancement of optical nonlinearities~\cite{Shan2021}.
A differential DC anomalous Hall conductivity was shown in \cite{McIver2019}.

Here we demonstrate a new method for generating and probing these phenomena by harmonically related optical fields which we call Harmonic Floquet Spectroscopy.
We utilize graphene, a well understood topologically trivial semi-metal, with an equilibrium band structure that lacks Berry curvature.
We generate our FTI by light dressing graphene, and subsequently control electrons using a second harmonic optical field within this new light-dressed band structure generating photocurrents.
Thus, we probe topological physics in a vastly different region of the dressed band structure than previously explored~\cite{McIver2019}.
Intriguingly, strongly dressing the band structure opens topological band gaps at harmonics of the dressing energy, allowing us to probe topological physics in a unique way~\cite{Kitagawa2010}.
Our novel approach enables control and simultaneous analysis of topological and sub-optical-cycle (attosecond) phenomena, combining the traditionally static picture of a dressed band structure with the sub-optical-cycle micromotion of the FTI state.
Here we accomplish this and pave the way to Floquet topotronics and its all-optical implementation.

\vspace{0.5cm}
\noindent
\textbf{FTI state generation}

\begin{figure*}[h]
	\includegraphics[width=\linewidth]{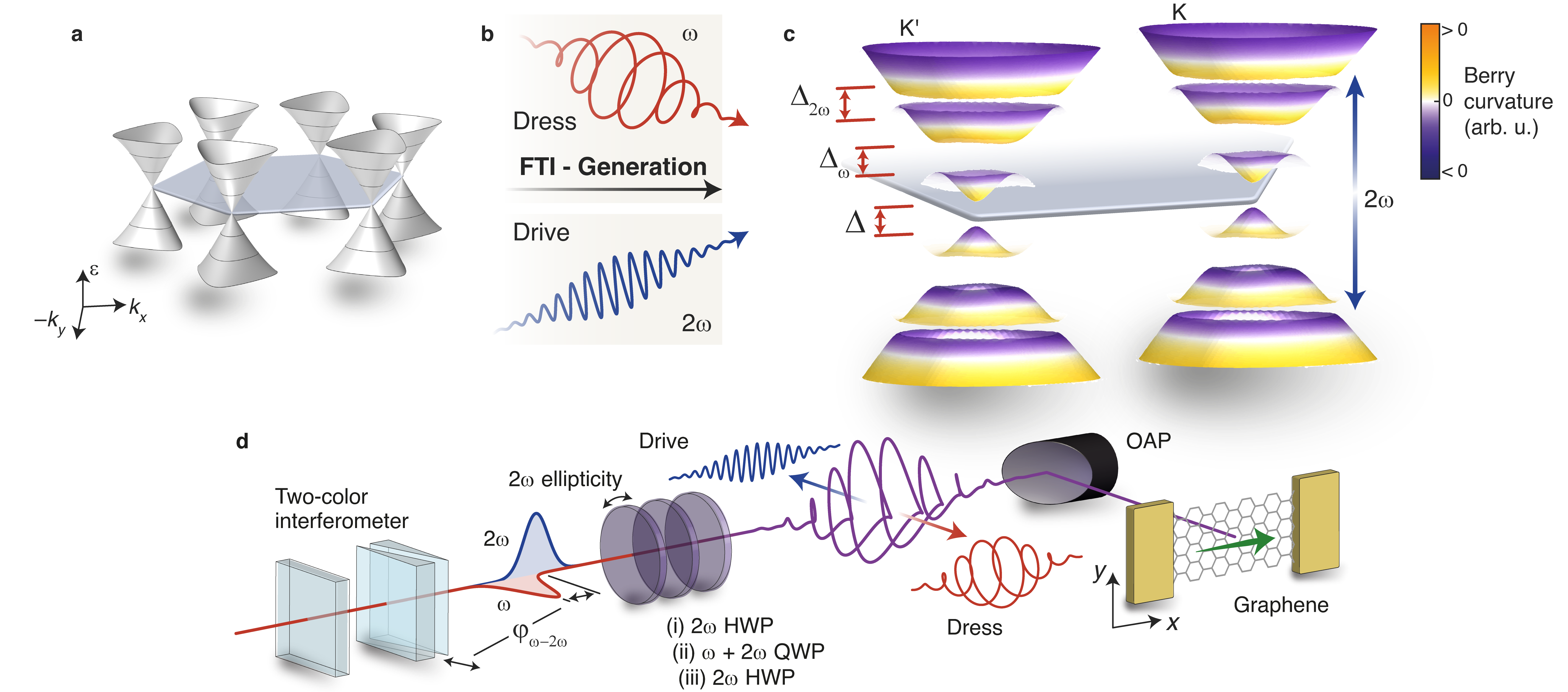}
    \caption{\textbf{Lightwave control of electrons in a Floquet topological insulator.} \textbf{a--c}, A Floquet topological insulator (FTI) emerges from driving the gapless, topologically trivial band structure of graphene (\textbf{a}) by irradiation  with a circularly-polarized fundamental laser field (\textbf{b}, red waveform). This dressing breaks only time-reversal symmetry, giving rise to both a nonzero gap $\Delta$ at the emerging K and K' valleys and a Berry curvature (colorbar) of the same sign in both valleys, resulting in a non-zero Chern number and a non-trivial topological insulating phase, with avoided crossings at resonant energies ($\Delta_{n\omega}$) (\textbf{c}, see Fig.~S2 for details).
    A second harmonic pulse (\textbf{b}, blue waveform) controls the motion of electrons in the FTI bands. 
    We probe circular dichroism and the anomalous Hall effect via photocurrents as signatures of the Berry curvature in the Floquet insulating phase. \textbf{d}, Sketch of the experimental setup. The phase \pr\ between the co-propagating $\omega$ and $2\omega$ pulses is controlled in a collinear two-color interferometer. In a sequence of dichroic half-waveplates (HWP) and a superachromatic quarter-waveplate (QWP; see Methods) we shape the $\omega$ circular dress and the $2\omega$ drive pulse with arbitrary polarization. Using an off-axis parabolic mirror (OAP), we focus the resulting waveform on a monolayer graphene strip, mounted in a vacuum chamber (not shown). We measure \pr- and polarization-dependent photocurrents via gold electrodes attached to the graphene strip.
    }
    \label{Fig0}
\end{figure*}

To generate our FTI, we optically dress bare graphene (Fig.~\ref{Fig0}a) with a circularly polarized fundamental laser field (with carrier angular frequency $\omega$, Fig.~\ref{Fig0}b).
This opens topological bandgaps within the dressed band structure (Fig.~\ref{Fig0}c).
In this transient topological band structure, a second `drive' laser pulse (with carrier angular frequency $2\omega$, Fig.~\ref{Fig0}b) drives coherent ultrafast electron dynamics in the light-dressed state, generating photocurrents.

While traditional measurements of transport phenomena in Floquet topological insulators are done with DC or low frequency fields, here we use an all-optical approach, which will be critical for elucidating new phenomena not seen before in solid state FTIs.
Due to strong dressing, our FTI exhibits avoided crossings at harmonics of the dressing energy, each capable of topological phenomena~\cite{Kitagawa2010} (calculated band structure color coded with Berry curvature, $\mathbf{\Omega}(\mathbf{k})$, $\Delta_{2\omega}$, Fig.~\ref{Fig0}c).
The second harmonic predominately probes at the topological bandgap, $\Delta_{2\omega}$, determined by the optical frequency and resulting conduction band populations, see Supplemental Text. To the best of our knowledge, FTIs have not been probed in this optical region.

By optically strongly modifying the material with the dressing light and subsequently driving electrons with the second harmonic, we are able to perform two unique measurements.
First, the micromotion (the sub-optical-cycle motion of the Floquet state resulting in attosecond phenomena) is mapped directly onto the two-color-phase, \pr , the phase difference between $\omega$ and $2 \omega$ light (Fig.~\ref{Fig0}d)~\cite{micromotion2014,Goldman2014,Eckardt2015,Rudner2020,Weitenberg2021,microarpes_2024}.
Second, the helicity (i.e., symmetry) of the second harmonic directly probes the symmetry of the Floquet state~\cite{Goldman2014,Eckardt2015,Schler2020}, i.e., the Berry curvature $\mathbf{\Omega}(\mathbf{k})$ around the $2\omega$ resonance (Fig.~\ref{Fig0}c, right).
By being able to generate the FTI state with the fundamental, and independently vary the ellipticity and \pr\ of the second harmonic, we can directly visualize quantum phenomena arising from the transient light-dressed FTI state stroboscopically, taking full advantage of Harmonic Floquet Spectroscopy.

Experimentally, we measure optical waveform-dependent photocurrents resulting from the two-color setup for different ellipticities and relative phases. The optical setup used for the synthesis of the two fields is sketched in Fig.~\ref{Fig0}d. Part of the light from an {80}\ {MHz} femtosecond erbium fiber laser is frequency doubled to form $2\omega$ pulses with a photon energy of {1.6}~{eV} ({775}~{nm}, {110}~{fs} intensity full width at half maximum, FWHM), which we overlap in a highly stable collinear setup with the fundamental $\omega$ pulses at {0.8}~{eV} ({1550}~{nm}, {213}~{fs} FWHM). The relative phase \pr\ between the pulses is controlled by calcite wedges in the collinear two-color interferometer~\cite{Brida2012}. After polarization control with a sequence of dichroic waveplates, the pulses are focused onto epitaxial monolayer graphene on silicon carbide using an off-axis parabolic mirror. The resulting currents are measured via attached gold electrodes. The fundamental field strength at the sample focus is measured to be {0.27}~{V/nm}, with a 3:4 field ratio of second harmonic to fundamental (equivalently, a ratio of 3:8 in vector potential, see Methods).
The frequency choice for the dressing and drive pulses results in short optical cycles avoiding decoherence~\cite{Ito2023} and excitation well above the Fermi level of graphene on SiC ($\sim$300~meV)~\cite{McIver2019}, placing us in the high-frequency driving limit~\cite{delaTorre2021}. Specifically, the optical driving period of $\sim$\,5fs is faster than electronic decoherence time in graphene~\cite{Heide2021c}, allowing us to generate and probe nonequilibrium states before decoherence and thermal relaxation take hold~\cite{delaTorre2021}. 
If and how dissipation from graphene to SiC plays a role needs to remain to future research. 
Importantly, dressing with strong near-infrared fields still allows for relevant gap sizes to open (i.e., a 53 meV bandgap at K/K', see Supplemental Text for more information) while taking advantage of mature, high-power, and low-noise optical sources~\cite{Lesko2021}. This results in a Floquet parameter\cite{Shan2021} $\mathcal{F} = \frac{e a \mathrm{E}_{\omega}}{\hbar \omega}$ of 0.08 (with $e$ the electron charge, $a$ the lattice constant, $\mathrm{E}_\omega$ the dressing field strength, and $\hbar$ the reduced Planck’s constant). Our Floquet parameter, gap sizes, and electric field (and vector potential) suggest that we are within the strong field regime~\cite{Sato2019}. We note that graphene sits at the edge of the Haldane phase diagram, so any circular dressing renders the system topological. 

\begin{figure*}
\includegraphics[width=\linewidth]{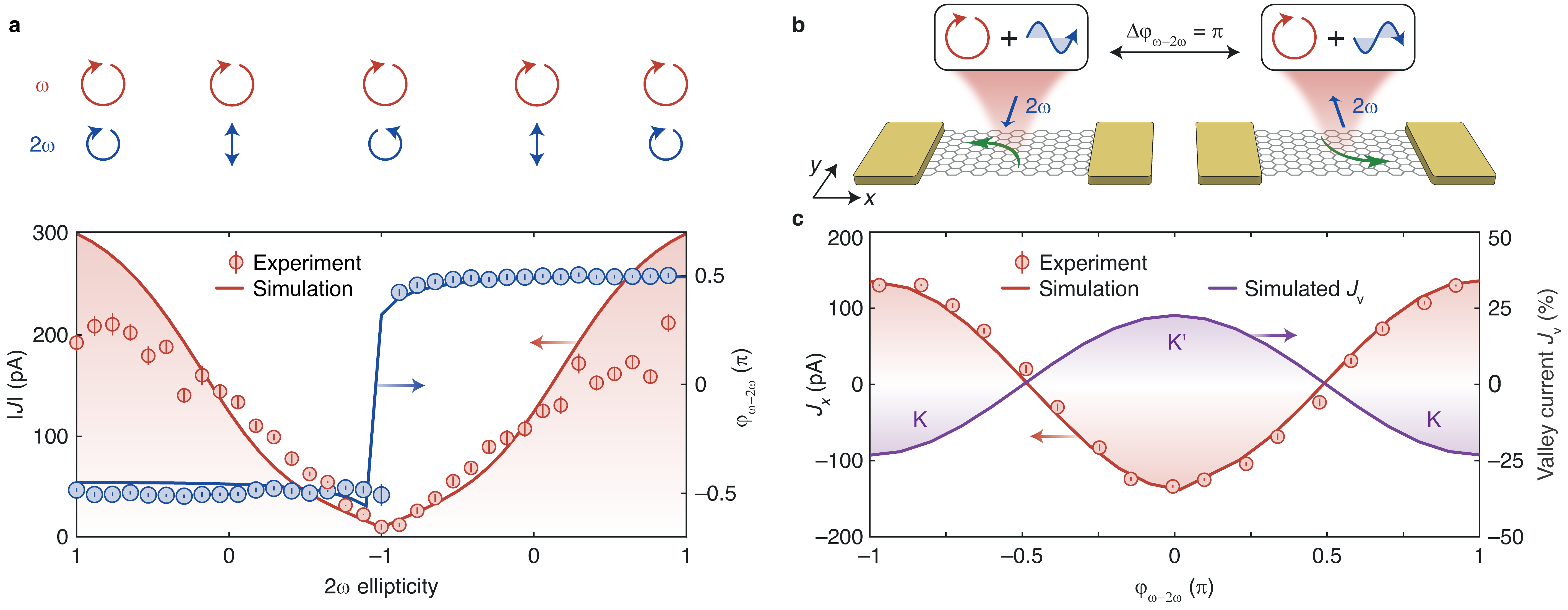}
\caption{
    \textbf{Circular dichroism and all-optical anomalous Hall effect.} 
    \textbf{a}, We probe circular dichroism by keeping the helicity of $\omega$ constant, generating the FTI, and continuously sweep the ellipticity of $2\omega$ (above). 
    Below, measured current amplitude $|J|$ (red circles) and phase $\theta$ (blue circles) as a function of the $2\omega$ ellipticity.
    TDDFT simulation results (lines) support the dichroic generation of ballistic photocurrent for co-rotating helicities of $\omega$ and $2\omega$ and suppression for counter-rotating helicities. 
    \textbf{b}, All-optical anomalous Hall currents (green arrows) emerge from excitation of the $\omega$-driven FTI with $2\omega$ control polarized perpendicularly to the electrode axis. Flipping \pr\ by $\pi$ effectively results in a reversal of the $2\omega$ deflection and consequently also the Hall current direction. 
    \textbf{c}, Measured current (red circles) as a function of \pr. 
    The computed current density (red line) confirms the directional control of the anomalous Hall effect. 
    The computed degree of valley polarized current $J_\mathrm{V}$ (purple line) reaches nearly {25}{\%}, realizing preferential current generation from either K or K'. 
    For the \textit{ab-initio} calculations, pulses at {1550}~{nm} with $E_{0}={0.27}~{\text{V/nm}}$ are employed with {75}{\%} $2\omega$ field admixture. 
    Error bars indicate ten times the standard deviation.
    }
\label{Fig4}
\end{figure*}

\vspace{0.5cm}
\noindent
\textbf{Circular dichroism with an FTI}

We first analyze the circular dichroism (CD), i.e., the system's helicity-dependent response, and direct manifestation of the system's Berry curvature~\cite{Wang2013b}.
This is a probe of the broken symmetry (and topological property) of our FTI state (Fig.~\ref{Fig4}a).
Importantly, graphene naturally does not display a CD response due to both the inversion and time-reversal symmetry, resulting in zero Berry curvature.
Once graphene is dressed by the purely circular $\omega$ dressing field, the resulting FTI state only breaks time reversal symmetry~\cite{Oka2009,Schler2020}, enabling a CD response.
Thus, we directly probe the time reversal symmetry breaking phenomena because the excitation pathways are dependent on both the Berry curvature of the FTI state (arising only from broken time reversal symmetry) and the excitation light ($2\omega$) helicity.
Due to the $2\omega$ pulse's energy, we specifically sample the Berry curvature (imparted by the $\omega$ pulse helicity) around the second avoided crossing ($\Delta_{2\omega}$, Fig.~\ref{Fig0}c). This avoided crossing exhibits the same topological phenomena as the opened gap at the Dirac point~\cite{Kitagawa2010}.
When the helicity of the $2\omega$ probe is matched to the Berry curvature of the FTI band structure, strong photocurrents are generated (momentum imbalance of conduction band population); and when the $2\omega$ helicity is anti-matched to the Berry curvature, a strong suppression of photocurrents occurs~\cite{Wang2013,Wang2013b,Schler2020}.

We experimentally observe this photocurrent suppression by nearly two orders of magnitude as the $2\omega$ helicity goes from $\epsilon = 1$ to $\epsilon = -1$, while the $\omega$ helicity is fixed to $\epsilon = 1$, preserving the sign of $\mathbf{\Omega}$ (Fig.~\ref{Fig0}c). 
These results are well matched by \textit{ab-initio} time-dependent density functional theory simulations (TDDFT, full line in Fig.~\ref{Fig4}a, see Methods). 
We stress that the circular dichroism is obtained by reversing the helicity of the drive ($2\omega$), and not that of the dressing ($\omega$) field. 
This is fundamentally different from pump circular dichroism exhibited in inversion-asymmetric materials, where the helicity of the dressing ($\omega$) light couples to only one of the valleys in a gapped system~\cite{Mak2012,Yin2022}.
Hence, our work represents the first measurement of the FTI state's intrinsic non-trivial dichroic response, directly enabled by the optical approach we employ.

\vspace{0.5cm}
\noindent
\textbf{All-optical anomalous Hall effect in an FTI}

Next, we investigate the ultrafast motion of electrons driven by a now linearly-polarized drive pulse in the FTI band structure.
A key hallmark of the FTI phase is the appearance of the anomalous Hall effect~\cite{Oka2009,McIver2019,Sato2019} (schematically shown in Fig.~\ref{Fig4}b), where a current is measured orthogonally to an applied voltage. 
This effect stems from the same-signed Berry curvature present for population in the conduction bands for both K/K' valleys causing a deflection of electrons driven linearly by the second harmonic, $\mathbf{j}=-\rho^\mathrm{c}(e/\hbar)\mathbf{E}_{2\omega}\times\mathbf{\Omega}$ (with excited population $\rho^\mathrm{c}$ and second harmonic driving field $\mathbf{E}_{2\omega}$).
Here, we orient the linear polarization of the second harmonic in $y$-direction and measure \pr-dependent currents along $x$ (Fig.~\ref{Fig4}b, c).
In a complete mapping of the photocurrent as a function of the two-color phase (Fig.~\ref{Fig4}c), we observe a reversal of the current between \pr$ = 0$ and $\pi$.
This shows that the optical phase controls the anomalous Hall current direction, and maps the micromotion of the FTI state~\cite{Goldman2014,micromotion2014} by the taking advantage of the shared periodicity of the fields.
We therefore demonstrate an attosecond measurement in a real solid, analogous to microsecond motion in synthetic cold atom lattice systems~\cite{Desbuquois2017,Wintersperger2020}.
Most importantly, this represents the first all-optically induced anomalous Hall effect measurement, in graphene, coherently controlled by two-color attosecond sub-cycle optical delay.

\textit{Ab-initio} TDDFT simulations that do not include edge states (Fig.~\ref{Fig4}a, c) agree remarkably well with the experiment, see Methods, implying that we can rule out the contribution of edge states in this geometry.
Further analysis of TDDFT results suggests that the primary mechanism is from Floquet physics, due to the majority role of the first valence and conduction bands of graphene, the alignment of population with dressed bands, and the lack of observed electron-electron interactions, see Methods.

Interestingly, the excellent agreement of experiment and theory allows us to unravel the origin of the currents: A decomposition of the current into its contributions from the K and K' valley shows that it is $\sim${25}{\%} valley selective (Fig.~\ref{Fig4}c purple line, 31 pA maximum), i.e.,\ one valley carries a larger population imbalance, see Methods. 
Despite the energetically indistinguishable K and K’ valleys, broken time reversal symmetry in the FTI valleys allows graphene to be harnessed for valleytronics, establishing FTI-based valleytronics.

\begin{figure*}[h!]
	\includegraphics[width=\linewidth]{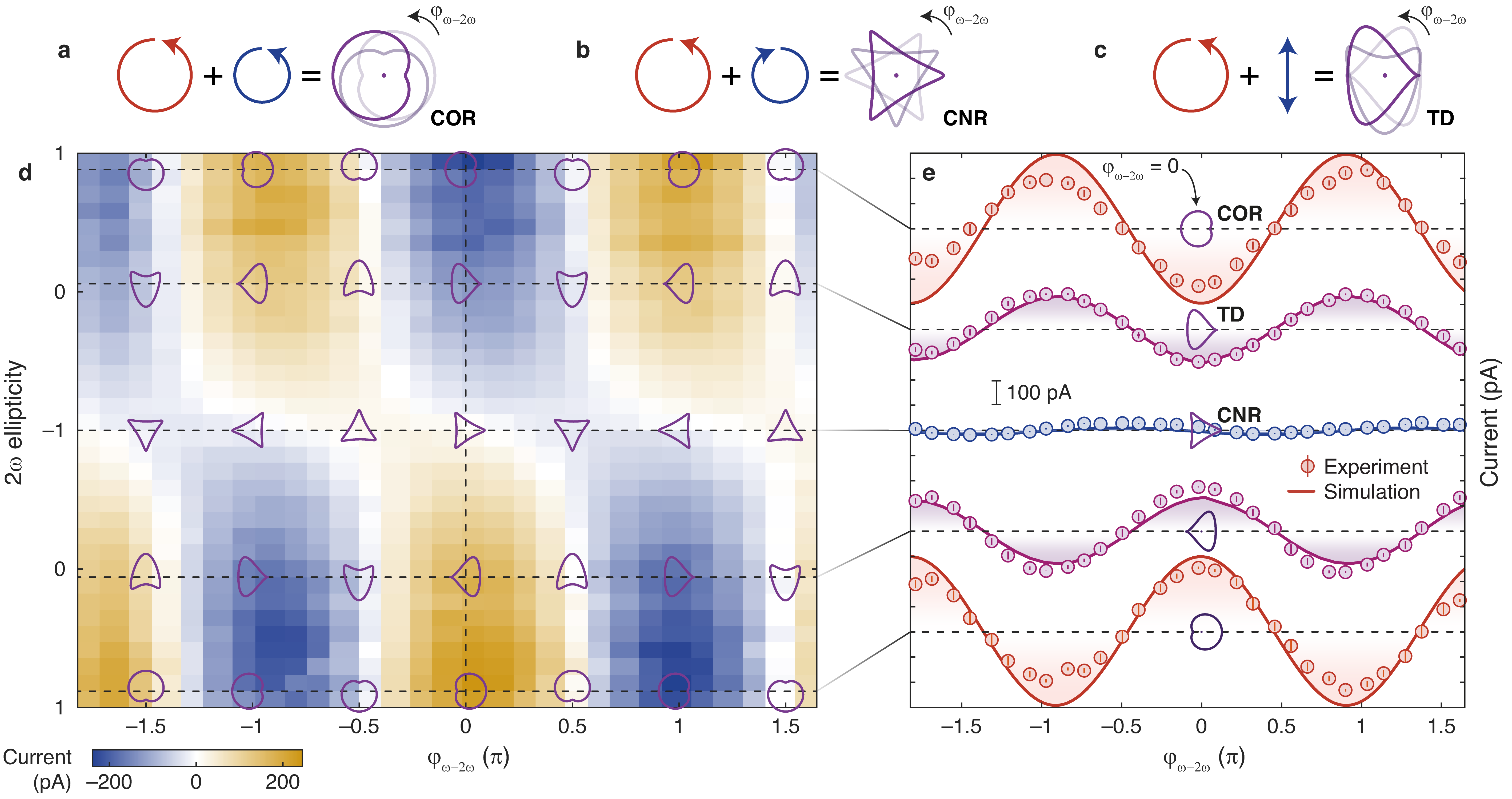}
	\caption{\textbf{Relative phase and polarization control over measured photocurrents.} 
    For clarity, we highlight three examples of the superposition of $\omega+2\omega$ fields (Lissajous figures) resulting in co-rotating circularly-polarized (\textbf{a}, COR), counter-rotating circularly-polarized (\textbf{b}, CNR), and teardrop-polarized (\textbf{c}, TD) waveforms, with variable orientation determined by the two-color phase (\pr).
    \textbf{d}, 2D map of relative phase-dependent current as a function of \pr\ and the $2\omega$ ellipticity. The insets exemplify the Lissajous figure transformation. \textbf{e}, Lineouts (horizontal dashed lines in \textbf{d}) as a function of \pr\, indicating the generation of current (red, COR), valley current (purple, TD) and valley polarization (blue, CNR) in the absence of current. Circles represent experimental data with error bars of ten times the standard deviation, solid lines indicate results from TDDFT.}
\label{Fig2}
\end{figure*}

\vspace{0.5cm}
\noindent
\textbf{Optical electron control within the FTI}

We next explore photocurrent generation in the topologically non-trivial graphene FTI, measuring all combinations of drive ($2\omega$) ellipticity and \pr\, with a fixed dressing field (Fig.~\ref{Fig2}).
To aid in the clarity for both the evolving $2\omega$ ellipticity and \pr, we utilize Lissajous figures; three examples are shown in Figs.~\ref{Fig2}a--c. 
These Lissajous figures result from the superposition of the fundamental laser fields ($\omega$) and its second harmonic ($2 \omega$) with properly chosen polarization states and \pr\ (Fig.~\ref{Fig0}d).

With matching fundamental and second harmonic helicities, a co-rotating circularly-polarized (COR) waveform emerges (Fig.~\ref{Fig2}a).
As discussed previously, the matched symmetries (i.e. helicity of $2\omega$ and Berry curvature $\mathbf{\Omega}$) results in strong photocurrent generation.
Changing \pr\ results in the $2\omega$ drive field sampling a different FTI micromotion phase, resulting in a reversal of the photocurrent.
Similarly, the Lissajous figure exhibits a dynamical (i.e., time-periodic) symmetry plane in one direction that rotates upon changing \pr\ \cite{Neufeld2019}.
With opposite helicities of fundamental and second harmonic, a counter-rotating circularly-polarized (CNR) waveform emerges and exhibits a three-fold rotational dynamical symmetry (Fig.~\ref{Fig2}b)~\cite{Neufeld2019}. 
As shown in the circular dichroism measurement (Fig.~\ref{Fig4}a), this $2\omega$ helicity results in photocurrent suppression.
By tuning \pr\ the Lissajous figure rotates.
This waveform has been studied extensively for producing inter-valley population imbalances, i.e., {\it valley polarization}~\cite{Kundu2016,Mrudul2021,JimenezGalan2021,Mitra2024,Tyulnev2024}.
While our FTI state (with only broken time reversal symmetry) has energy-equivalent K/K' points, the non-equivalent orbital character of K/K' allow us to use the \pr\ to address them differently.
This FTI state is fundamentally different from systems with broken inversion symmetry where the K/K' points are energy-inequivalent.
Finally, the last case with a linear $2\omega$  drive field is shown in Figure~\ref{Fig2}c.
As \pr\ changes, the Lissajous figure continuously deforms and rotates the combined field's symmetry axis.
Revealed by \textit{ab-initio} simulations, this teardrop (TD) waveform leads to a current dominantly arising from one valley only, i.e., a {\it valley current.}

To experimentally investigate these three distinct cases, we map out photocurrents as a function of all relevant $2\omega$ polarization states and phases (Fig.~\ref{Fig2}d). 
On the vertical axis, as we tune the $2\omega$ ellipticity while keeping $\omega$ circular, we continuously sweep from a COR to a TD and a CNR waveform and further to a TD and COR waveform again. 
As the fast and slow axes of the $2\omega$ half waveplate are interchanged by sweeping across ellipticity $-1$, a $\pi$ phase jump is introduced between the two colors, resulting in an inversion of current direction with approximately identical amplitudes for equal ellipticities $>-1$. 
The remarkably large magnitude of up to {214}~{pA} obtained for the COR waveform can be assigned to the substantially more efficient momentum symmetry breaking compared to carrier-envelope phase-controlled current generation~\cite{Langer2020,Heide2021,Hanus2021,Boolakee2022}. 
As the waveplate is rotated to generate the TD waveform, $|J|$ decreases to approximately {140}~{pA}. 
Finally, with the CNR waveform, $|J|$ further decreases to a minimum of {12}~{pA}. 
Importantly, the strong overall suppression of current injection supports the generation of valley polarization by the bichromatic waveforms (see Supplemental Text).

Changing \pr\ with the two-color interferometer results in a continuous change of the Lissajous figures, which we observe experimentally as a sinusoidal oscillation of injected \textit{x}-direction current (Fig.~\ref{Fig2}d and lineouts, solid circles, in Fig.~\ref{Fig2}e). 
This oscillatory behavior not only maps the Floquet state micromotion~\cite{Eckardt2015,micromotion2014,Wintersperger2020,microarpes_2024}, but is also known from $\omega-2\omega$ coherent control~\cite{Atanasov1996,Shapiro2011,Eckardt2015}, which we will connect below with dynamical symmetry selection rules imposed by the excitation waveform~\cite{Neufeld2019,Neufeld2021}.
Overlaid on the experimental data is the calculated current from TDDFT without any free parameters (normalized to the TD waveform current amplitude, see Methods).
We observe similar sinusoidal behavior between the experiment and theory, as well as a small non-zero current for the CNR waveform with a phase-shifted maximum, showing remarkable agreement with the measurement.

\begin{figure*}[h!]
    \includegraphics[width=\linewidth]{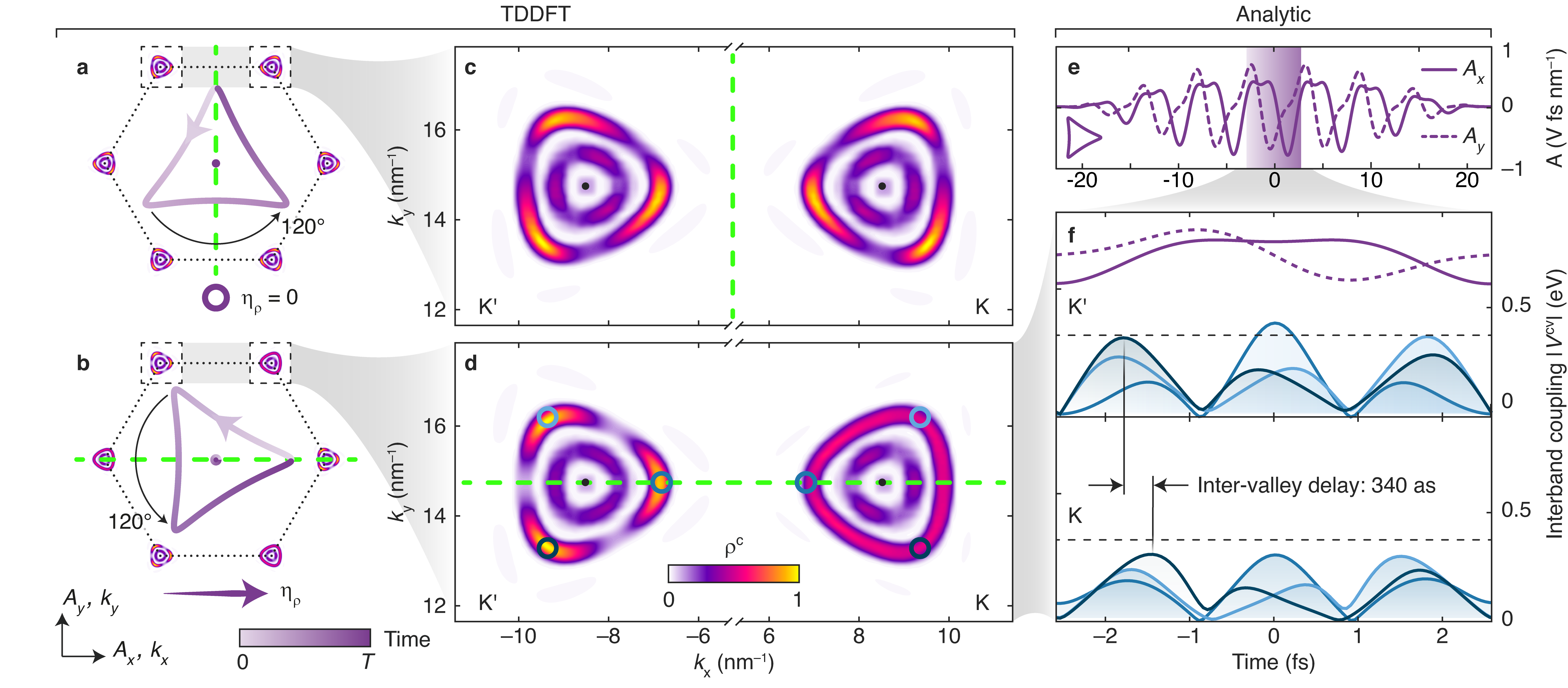}
    \caption{\textbf{Sub-cycle dynamical symmetry selection rules.} \textbf{a--b}, Temporal shape and symmetries of the vector potential Lissajous figures (purple lines) and the resulting symmetries of residual momentum distributions in the conduction band ($\rho^\mathrm{c}$) imprinted to the first Brillouin zone (dotted hexagons) for the CNR waveforms for $\varphi_\mathrm{\omega-2\omega}=\pi/2$ (a) and $0$ (b) respectively. The momentum distributions are computed with TDDFT, see Methods. 
    Green dashed lines indicate global symmetry axes, determined by both waveform and time periodicity (see purple color bar).
    \textbf{c--d}, Close-ups of $\rho^\mathrm{c}$ around K and K' as indicated in \textbf{a, b}. 
    For all cases, an eight-cycle pulse (vector potential shown in E) is applied with {1550}~{nm} fundamental wavelength with a peak field strength $E_{0}={0.27}\,{\text{V/nm}}$ and {75}{\%} $2\omega$ field admixture. 
    \textbf{e}, CNR vector potential waveform for $\varphi_{\omega-2\omega}=0$. 
    The shaded area indicates the center optical cycle. 
    \textbf{f}, temporal profile of the interband dipole coupling $|V^\mathrm{cv}|$~\cite{Weitz2024} induced by the waveforms shown in E. 
    The different shades of blue refer to initial momenta in the K and K' valley as marked with circles in panel D. The dashed black guide lines compare the $|V^\mathrm{cv}|$ amplitudes at K and K'.}
    \label{Fig3}
\end{figure*}

\vspace{0.5cm}
\noindent
\textbf{Symmetries, selection rules and sub-cycle dynamics}

For further insight, it is highly instructive to investigate how the $\omega-2\omega$ light field symmetries are reflected in the excited state populations.
Critically, these symmetries will allow us to understand when photocurrent generation is allowed or forbidden (but will never predict the magnitude and scaling, as part of the current will originate from the Berry curvature of the FTI state).
This analysis will connect the sub-optical-cycle electron motion in the {\it undressed} band structure to the topological phenomena and micromotion present in the FTI picture. 

Figures~\ref{Fig3}a--d show the residual conduction band distributions for the CNR waveform obtained from TDDFT, see Methods. 
Analyzing all six K and K' points (Figs.~\ref{Fig3}a--b) allows us to look at the dynamical (periodic) symmetry selection rules, while the zoom-ins (Figs.~\ref{Fig3}c--d) highlight the intra-valley symmetries~\cite{Neufeld2019,Neufeld2021}.
Two symmetries dominate: (i) a mirror axis coupled to time-reversal in all geometries (dashed green lines), and (ii) a three-fold rotational symmetry coupled to time translations in the CNR case only. 
These symmetry operators map onto the momentum distributions of electrons in the conduction band.
From these symmetries we can assign selection rules to how the light field symmetries dictate phenomena such as valley polarization and valley current, representing the first experimental mapping (Fig.~\ref{Fig2}d) of these dynamical symmetry selection rules.

We analyze these dynamical selection rules by looking at momentum distributions in Figure~\ref{Fig2}d where the residual measured current is experimentally zero (indicating a forbidden injection current). 
For the CNR waveforms (Figs.~\ref{Fig3}a, b), we see one mirror axis present in the electron excitation distributions; a M-$\Gamma$-M symmetry axis for \pr = $\pi/2$ (green dashed line, Fig.~\ref{Fig3}a), rendering the conduction band population ($\rho^\mathrm{c}$) K-K' symmetric, hindering the ability to generate valley polarization.
When \pr = $0$, we \textit{disconnect} the K and K' points rotating the symmetry axis to K-$\Gamma$-K' (green dashed line, Fig.~\ref{Fig3}b). 
This allows us address K/K' differently, and results in a non-zero valley polarization of $\eta_\rho={1}{\%}$, see Methods.
The zoom-ins of the CNR population (Figs.~\ref{Fig3}c, d) reveal the local $\rho^\mathrm{c}$ symmetries within the K and K' valleys. 
The total injection current should vanish due to the rotational dynamical symmetry in the CNR case (since current from all K and K' points individually cancel each other). 
Experimentally (and numerically in TDDFT) we observe a small residual current (Fig.~\ref{Fig2}e) that occurs from breaking of time-translational symmetry by the pulse envelope itself (i.e., departing from an infinitely time-periodic Floquet Hamiltonian). 
The COR and TD waveforms are further analyzed in the Supplemental Text.

To understand the emergence of the characteristic $\rho^\mathrm{c}$ patterns in Fig.~\ref{Fig3}d on a sub-cycle time scale, it is insightful to further discuss the time-dependent inter- and intraband electron dynamics in the {\it undressed} band structure.
We highlight these dynamics for three exemplary electrons with initial momenta equidistant to the respective K and K' points (Fig.~\ref{Fig3}d, colored circles). 
Figure~\ref{Fig3}f shows the respective sub-cycle evolution of the interband dipole coupling $V^\mathrm{cv}(t)=\mathbf{E}(t)\cdot\mathbf{d}^\mathrm{cv}$ (instantaneous Rabi frequency modulo $\hbar$), which approximates the instantaneous excitation probability for a given electric field $\mathbf{E}(t)$ and an interband transition dipole matrix element $\mathbf{d}^\textbf{cv}(\mathbf{k}(t))$~\cite{Weitz2024}. 
The stark inter-valley differences in $|V^\mathrm{cv}|$ occurring during the center optical cycle (Fig.~\ref{Fig3}e, shaded area) become clear with the corresponding trajectories drawn in Fig.~\ref{Fig3}f; The differences in the sub-femtosecond trajectories of electrons between K and K' show how the composite light field can be used to understand the valley-selective physics.
By looking at the coupling strength peaks, we observe inter-valley population differences forming within a time delay of 340\,attoseconds between two $k$-space mirrored points.
These dynamics are identical to $2\omega$-driven electrons sampling the micromotion of the $\omega$-induced topological FTI band structure.

\vspace{0.5cm}
\noindent
\textbf{Conclusions and outlook}

We have demonstrated photocurrents from a Floquet topological insulator, utilizing our novel spectroscopic probe Harmonic Floquet Spectroscopy, and thereby introducing photocurrents as a new probe for nonthermal excitations on ultrafast time scales. 
We measured the all-optical anomalous Hall effect and photocurrent circular dichroism utilizing our two-color optical setup.
The photocurrents show an exceedingly strong phase and ellipticity dependence, allowing us to use them to probe the ultrafast dynamics in the FTI. 
Excellent agreement with \textit{ab-initio} theory allowed us to pinpoint the source of currents to the FTI valleys at K and K'. 
This allow us to realize valley current (with a notable magnitude of 31\,pA) from a Floquet topological insulator, a decades long goal for valleytronics.
For a specific case, we considered the interaction of the two-color light field with the bare material, revealing experimentally measured lightwave dynamical symmetry photocurrent selection rules. 
This symmetry picture maps the sub-optical-cycle electron motion onto the topological photocurrents, showing the deep connection between these two distinct bases, the lightwave symmetries and the FTI states.

Our work heralds Floquet lightwave topotronics~\cite{Gilbert2021}, where topology is designed by an optical dressing field, almost unconstrained by material properties, and electrons are driven by a second field in the resulting FTI band structure, on (sub-) femtosecond time scales.
We do this by experimentally showcasing this through photocurrent circular-dichroism and the all-optical anomalous Hall effect, showing sub-optical-cycle control of these photocurrents and relating them to the attosecond micromotion inside the FTI. Future work will focus on understanding the role of dissipation and other material-related properties in extending functionalities of quantum dynamics far from equilibrium. 
We foresee direct ramifications in the realm of ultrafast quantum material science~\cite{delaTorre2021}, lightwave electronics, topological quantum computing, sensing, and ultrafast spectroscopy, even in conjunction with twistronics.

% \clearpage % Clear all remaining figures and tables then start a new page

% \bibliography{references.bib} %1-46, 47-56

%%%%%%%%%%%%%%%% Methods %%%%%%%%%%%%%%%
\section*{Methods}
\subsection*{Experimental Details}
We use laser pulses at a fundamental photon energy $\hbar\omega={0.8}$~eV ({1550}~{nm} central wavelength) with a pulse duration of {213}~{fs} (intensity full width at half maximum, FWHM) from an amplified 80~MHz erbium-doped fiber oscillator. After frequency doubling in a 1~mm thin type-I phase-matched bismuth borate crystal (BiBO), fundamental and second harmonic (photon energy $2\hbar\omega={1.6}$~eV, {775}~{nm} central wavelength, {110}~{fs} FWHM) pulse pairs co-propagate with orthogonal linear polarization (Fig.~\ref{Fig0}d, red and blue envelopes). 

A collinear two-color interferometer based on birefringent calcite elements is used to control their relative phase \pr\ (Fig.~\ref{Fig0}d)~\cite{Brida2012}. The first $d_\mathrm{a}=3$~mm thin calcite plate is oriented with its ordinary (o) axis parallel to the $2\omega$ polarization and its extraordinary (e) axis parallel to $\omega$; for the wedged calcite elements, the axes are rotated by 90$^{\circ}$ to invert the situation. As the two colors experience different refractive indices along the respective axes ($n_\omega^\mathrm{e}=1.47$, $n_\omega^\mathrm{o}=1.64$, $n_{2\omega}^\mathrm{e}=1.48$, $n_{2\omega}^\mathrm{o}=1.65$) this configuration allows us to provide the $\omega$ pulses with a head-start after the first calcite plate, which are subsequently caught up by the $2\omega$ pulses in the calcite wedge pair~\cite{Ghosh1999}. By finely tuning the inserted thickness $d_\mathrm{b}$ of the wedges, the relative phase is controlled as
\begin{equation}\label{eq:phi_rel}
    \varphi_{\omega-2\omega}=\frac{2\omega}{c}\left(d_\mathrm{a}\,\Delta n_{\omega-2\omega}^\mathrm{oe}+d_\mathrm{b}\,\Delta n_{\omega-2\omega}^\mathrm{eo}\right),
\end{equation}
with $\Delta n_{\omega-2\omega}^{kl}=n_\omega^k-n_{2\omega}^l$. Temporal overlap is determined by maximizing the $\omega+2\omega$ sum frequency signal obtained from a type-II phase-matched beta barium borate (BBO) crystal in a subsequent reference path. The collinear interferometer design enables a particularly high passive phase stability exhibiting {14.6}~{mrad} integrated jitter in the angular stability of the bichromatic Lissajous figures over 1~hour (see fig.~\ref{FigS1}). Pulse dispersion upon wedge insertion is negligible as we sample less than four periods of \pr\ with rather narrow-band spectra at both colors.

For individual control over the ellipticity and rotation of the polarization ellipse of both colors, the delayed pulses are subsequently sent through a sequence of (i) a dichroic half-waveplate (Newlight Photonics) acting on $2\omega$ only, (ii) a superachromatic quarter-waveplate (Thorlabs) acting on both colors, and (iii) a dichroic half-waveplate (Newlight Photonics, same as i), which counteracts the ellipsoidal tilt of the $2\omega$ polarization introduced by waveplate (ii) upon rotation of waveplate (i). As waveplate (i) is rotated (Fig.~\ref{Fig0}d), the ellipticity of the $2\omega$ polarization is changed from $+1$ (half-waveplate angle {45}$^\circ$) to $0$ (half-waveplate angle {22.5}{$^\circ$}) and finally $-1$ (half-waveplate angle {0}{$^\circ$}). Then it is changed back to $0$ (half-waveplate angle $-22.5^\circ$) and finally $+1$ (half-waveplate angle {$-45^\circ$}), while the $\omega$ polarization is kept constant at circular (ellipticity $+1$). This sequentially generates the COR, TD and CNR waveforms, and further the TD and COR waveforms with a phase slip of $\Delta\varphi_{\omega-2\omega}=\pi$. We carefully calibrated the waveplates' retardation to a wire-grid polarizer and ensured their dichroic performance with combined long- and shortpass filters. We measure a polarization extinction ratio of 14 dB for the combined harmonics, resulting in an ellipticity of 0.96. The resulting bichromatic waveforms are tightly focused onto a sample using a 15~mm off-axis parabolic mirror (Fig.~\ref{Fig0}d).

We use a $2\times{10}$~$\mu m^2$ strip of epitaxially grown monolayer graphene on silicon carbide as a sample. Its short sides are connected to {30}~{nm} gold electrodes with a 5~nm titanium adhesion layer. Details on the fabrication and characterization are provided in~\cite{Boolakee2022}. Gold bond wires connect the electrodes with a chip carrier for photocurrent measurements. All measurements are performed under high vacuum conditions ({$10^{-8}$}~{hPa}) at room temperature.

For photocurrent detection, we focus the pulse train to the center of the structure such that only the graphene is illuminated. To isolate waveform-dependent currents from photo-thermoelectric currents and other static background, we modulate \pr\ from pulse to pulse at $f_\mathrm{mod}={117}$~{Hz} with a piezoelectric actuator mounted to one of the calcite wedges. We amplify the resulting current via transimpedance amplification and record it via dual-phase lock-in detection referenced to $f_\mathrm{mod}$. Each data point shown in Fig.~\ref{Fig2} is recorded with a 300-ms time constant and averaged over 25 data samples. The amplitude of the piezo stroke is calibrated via an external laser diode Michelson interferometer with one end mirror attached to the modulated calcite wedge. When the resulting modulation depth of \pr\ reaches $\pi$, the signal amplitude detected by the lock-in amplifier is maximized. The measured lock-in phase, $\theta$ (Fig.~\ref{Fig4}b), is equivalent to the absolute phase \pr\, as we determined the temporal overlap of both colors.

\subsection*{Floquet Calculations}
We employ a Floquet Hamiltonian for the light-dressed graphene from a two-band next-neighbor tight-binding Hamiltonian
\begin{equation}\label{eq:NN_ham}
	\mathcal{H}_0(\mathbf{k})=\gamma\begin{bmatrix}
		0 & f(\mathbf{k}) \\ f^\ast(\mathbf{k}) & 0\end{bmatrix},
\end{equation}
with a hopping parameter $\gamma={-2.9}$~eV and 
\begin{widetext}
\begin{equation}\label{eq:fk}
    f(\mathbf{k})=\exp\left(i\frac{ak_y}{\sqrt{3}}\right)+2\exp\left(-i\frac{ak_y}{2\sqrt{3}}\right)\cos\left(\frac{ak_x}{2}\right),
\end{equation}
\end{widetext}
where $a={2.46}$~\r{A} is the lattice constant of graphene. By coupling $\mathcal{H}_0(\mathbf{k})$ to the desired continuous-waveform $\mathbf{E}(t)$ of fundamental periodicity $T_0=\tfrac{2\pi}{\omega}$ as $\mathbf{k}(t)=\mathbf{k}_0+\tfrac{e}{\hbar}\int_{-\infty}^{t}\mathbf{E}(t')\mathrm{d}t'$, we obtain the sub-blocks of the Floquet Hamiltonian as~\cite{Giovannini2019}
\begin{widetext}
\begin{equation}\label{eq:Floquet_ham}
	\mathcal{H}_\mathrm{F}^{mn}(\mathbf{k}_0)=\frac{1}{T_0}\int_{0}^{T_0}\mathcal{H}_0(\mathbf{k}(t))\mathrm{e}^{i(m-n)\omega t}\mathrm{d}t+\delta_{mn}m\hbar\omega\sigma_0.
\end{equation}
\end{widetext}
$|m-n|$ is the order of the Floquet replica, which we numerically expand up to order 5. The replica are obtained by diagonalizing $\mathcal{H}_\mathrm{F}$.
We plot the resulting Floquet band structure in fig.~\ref{FigFloquet}a, where the color weight is the overlap with the undressed band structure~\cite{Oka2009,Sato2019}. The calculated density of states (fig.~\ref{FigFloquet}b) shows open bandgaps at the K point, as well as the one ($\omega$) and two photon ($2\omega$) resonances. A quantized Hall effect requires a gap in the density of states which can be possibly realized in the two-color experiment.

We directly calculate the Berry curvature as a summation of the eigenstates of the Floquet Hamiltonian,
\begin{widetext}
\begin{equation}\label{eq:BerryCurve}
    \Omega_{\mu,\nu}^{\mathrm{n}} (\mathbf{R}) = i \sum_{\mathrm{n'} \neq \mathrm{n}} \frac{\langle \mathrm{n} | \partial \mathcal{H}_\mathrm{F}/\partial \mathrm{R}^{\mu}| \mathrm{n'} \rangle \langle \mathrm{n'} | \partial \mathcal{H}_\mathrm{F}/\partial \mathrm{R}^{\nu}| \mathrm{n} \rangle - (\nu \leftrightarrow \mu)}{(\epsilon_{\mathrm{n}}  - \epsilon_{\mathrm{n'}})^2} 
\end{equation}
\end{widetext}
due to the ability to avoid differentiation giving a freedom of gauge choice~\cite{Xiao2010}.
We plot the band structure in figures~\ref{FigFloquet}c--f, where the color encodes the curvature.
Figures~\ref{FigFloquet}d--f show the topological bandgap of 53~meV opened at the K point (d), 132~meV gap at the one photon resonance (e, 10x Berry curvature), and 27~meV gap at the two photon resonance (f, 10x Berry curvature). 
It is evident that the circular driving imparts Berry curvature at the relevant avoided crossings of the band structure.
The role of this added Berry curvature can be explored by scaling the bandgap size.

Finally, we can simulate the Floquet states (fig.~\ref{FigFloquetScaling}) for experimentally achievable field strengths at 1550~nm~\cite{Lesko2021,Lesko2022}. Importantly, the shorter wavelength dressing fields allows for significantly larger bandgaps (100s of meV) in graphene when compared to longer driving laser wavelengths. It is worth noting that these field strengths have already been used on graphene in previous work at 800~nm~\cite{Weitz2024}.

\subsection*{Ab-initio TDDFT simulations}
We describe here details of the \textit{ab-initio} calculations employed throughout the paper. All calculations are based on TDDFT simulations performed with the open access code Octopus~\cite{TancogneDejean2020}. We solved the time dependent Kohn-Sham (KS) equations of motion, given in the velocity gauge and atomic units by:
\begin{widetext}
\begin{equation}
\label{eq:kseom}
	i\partial_{t}\psi_{n,\textbf{k}}(\mathbf{r},t) =	
        (\frac{1}{2}(-i\mathbf{\nabla} + \textbf{A}(t)/c)^2 + v_\mathrm{KS}(\mathbf{r},t))\psi_{n,\textbf{k}}(\mathbf{r},t),
\end{equation}
\end{widetext}
where $\psi_{n,\textbf{k}}(\mathbf{r},t)$ is the KS-Bloch (KSB) state at band index $n$ and \textit{k}-point $\textbf{k}$, and $v_\mathrm{KS}$ is the KS potential that comprises the classical Hartree interaction term, the interactions of electrons with nuclei and deeper electronic states (which were described by norm-conserving pseudopotentials~\cite{Hartwigsen1998}), and the exchange-correlation (XC) term, where we employed the adiabatic local density approximation (aLDA). We have assumed partial periodic boundary conditions in the graphene monolayer plane (\textit{xy}), while the \textit{z}-axis was treated with finite boundary conditions and a complex absorber of width 12 Bohr during propagation (similar to ref.~\cite{Neufeld2021}). In eq.~\ref{eq:kseom}, $\mathbf{A}(t)$ is the applied vector potential after assuming the dipole approximation, which was taken to have the following form:
\begin{widetext}
\begin{equation}
\label{eq:vector}
\mathbf{A}(t)=f(t)cE_{0}(\frac{1}{\omega \sqrt{2}}(\cos(\omega t+\varphi_{\omega-2\omega})\hat{\mathbf{x}}+\sin(\omega t+\varphi_{\omega-2\omega})\hat{\mathbf{y}}) + \frac{1}{2\omega \sqrt{1+\epsilon^2}}(\epsilon\cos(2\omega t)\hat{\mathbf{x}}+\sin(2\omega t)\hat{\mathbf{y}}))
\end{equation}
\end{widetext}
with $c$ the speed of light, $E_{0}$ the electric field amplitude (taken at experimental values), $\varphi_{\omega-2\omega}$ the two-color phase, $\epsilon$ the ellipticity of the $2\omega$ field, $\omega$ the carrier frequency (taken at the experimental value), and $f(t)$ a temporal envelope (called a `super-sine') taken as~\cite{supsin}: 
\begin{equation}
\label{eq:env}
f(t) = \sin{\left(\pi \frac{t}{T_\mathrm{p}}\right)}^{\left(\frac{|\pi(\frac{t}{T_\mathrm{p}}-\frac{1}{2})|}{\sigma}\right)},
\end{equation}
where $\sigma=0.75$, $T_\mathrm{p}$ is the duration of the laser pulse, which was taken to be $T_\mathrm{p}=8T$, and $T={5.17}$~{fs} is a single cycle of the fundamental carrier frequency. This leads to an applied electric field of duration 20.6~fs (FWHM). In eq.~\ref{eq:vector} by varying the relative phase and ellipticity we obtain the waveforms plotted in the main text.

From the propagation of the KSB states we obtained the current expectation value: 
\begin{widetext}
\begin{equation}
\label{eq:curr}
\textbf{J}(t) = \sum_{n,k}w_{k}\int [\psi^{\dag}_{n,\textbf{k}}(\mathbf{r},t)(\frac{1}{2}(-i\mathbf{\nabla} + \textbf{A}(t)/c) - i[V_\mathrm{ion},\textbf{r}])\psi_{n,\textbf{k}}(\mathbf{r},t)]\mathrm{d}\textbf{r}  + c.c.,
\end{equation}
\end{widetext}
where $V_\mathrm{ion}$ is the pseudopotential non-local part of $v_\mathrm{KS}$, $w_k$ is the \textit{k}-point weight, and the sum is performed over occupied states. From $\textbf{J}(t)$, we calculated the actual measured current by averaging $\textbf{J}(t)$ over a single cycle of the fundamental frequency after the driving laser pulse has ended (at $t=t_f$): $\textbf{j}=\int_{t_f}^{t_f+T} \textbf{J}(t) \mathrm{d}t$.
We calculated the projection plots in Fig.~\ref{Fig3} by projecting the time-dependent KSB states after the laser pulse has ended onto the ground states: 
\begin{equation}
    \label{eq:proj}
    g_{n}(\textbf{k}) = \sum_{m}|\int \psi^{\dag}_{n,\textbf{k}}(\mathbf{r},t)\psi_{m,\textbf{k}}(\mathbf{r},t)\mathrm{d}\textbf{r}|^2,
\end{equation}
where the sum runs over occupied states. From $g_{n}(\textbf{k})$ we obtained band-resolved occupations, or summed over all conduction/valence bands as presented in Fig.~\ref{Fig3}. The final occupation plot was also interpolated on a denser \textit{k}-grid and filtered.

The \textit{ab-initio} calculated phase in Fig.~\ref{Fig4}b was obtained from the injection current's direction in the \textit{xy} plane. The valley current in Fig.~\ref{Fig4}d was calculated by integrating separately for $k_x>0$ and $k_x<0$ regions in \textit{k}-space in eq.~\ref{eq:proj} for $\textbf{J}(t)$, and using those quantities to define the normalized contribution of each half of the BZ:
\begin{equation}\label{eq:eta}
J_{\mathrm{V}}=\frac{|\textbf{j}_\mathrm{K}|-|\textbf{j}_\mathrm{K'}|}{|\textbf{j}_\mathrm{K}|+|\textbf{j}_\mathrm{K'}|},
\end{equation}
with $|\textbf{j}_\mathrm{K/K'}|$ being the injection current contributed from each region in the BZ. 

We performed further auxiliary calculations where the time dependent KS potential was frozen to its ground state form at $t=0$, i.e., $v_\mathrm{KS}(\textbf{r},t)=v_\mathrm{KS}(\textbf{r},t=0)$. This yields the independent particle approximation (IPA), where the different KSB states are decoupled from one another in the equations of motion and electron-electron interactions do not evolve dynamically. Figure~\ref{FigS2} compares the calculated current from the full TDDFT or IPA simulations for laser parameters similar to the experiment, but taken at an even higher laser power to try and induce larger correlations. The calculated current in both cases agrees remarkably well, indicating that electronic interactions do not play any role in the main mechanisms observed and discussed in the main text.

We further computed the relative occupations of the different valence and conduction bands after the laser pulse ends by integrating $g_{n}(\textbf{k})$ from eq.~\ref{eq:proj} over \textit{k}-space. From this analysis we find that in typical experimental conditions, over 99.999$\%$ of the excitations occur in the first valence and first conduction bands. This result, together with the absence of dynamical interactions shown in fig.~\ref{FigS2}, validates the model two-band tight-binding Hamiltonian employed in the analytic approach~\cite{Weitz2024}.

%%%%%%%%%%%%%%%% ACKNOWLEDGEMENTS %%%%%%%%%%%%%%%

\subsection*{Acknowledgments:}
We thank Sebastian Lotter and Heiko B.\ Weber for providing the sample, as well as Monika Aidelsburger, Anna Galler, Gregor Jotzu, and Jacob Pettine for helpful discussions.
\subsection*{Funding:}
This work has been funded by the Deutsche Forschungsgemeinschaft (SFB 953 ‘Synthetic Carbon Allotropes’, 182849149), the PETACom project financed by Future and Emerging Technologies Open H2020 program, ERC Grant AccelOnChip (884217), and the Gordon and Betty Moore Foundation (GBMF11473).
C.H. acknowledges support from the Alexander von Humboldt Research Fellowship and US Department of Energy, Office of Science, Basic Energy Sciences, Chemical Sciences, Geosciences, and Biosciences Division through the AMOS program.
\subsection*{Author contributions:}
P.H. and T.W. conceived the study. T.W. and S.W. performed the experiment with input from C.H.. T.W. and D.M.B.L. analyzed the results. O.N., D.M.B.L, T.W. and W.L. performed the simulations. D.M.B.L., T.W., O.N. and P.H. wrote the manuscript with input from all authors. 
\subsection*{Competing interests:}
A subset of the authors have filed a patent titled ``Method and apparatus for generating valley current in a solid material" (EP24175546.1).
\subsection*{Data and materials availability:}
Source data are provided with this paper. All other data that support the plots within this Article and other findings of this study are available from the corresponding author(s) upon reasonable request.

%%%%%%%%%%%%%%%% SUPPLEMENT LIST %%%%%%%%%%%%%%%

% \noindent\subsection*{Supplementary materials}
% \noindent Materials and Methods\\
% Supplementary Text\\
% Figs. S1 to S10\\
% References \textit{(46-54)}\\ 

%%%%%%%%%%%%%%%% END OF MAIN TEXT %%%%%%%%%%%%%%%

\newpage

%%%%%%%%%%%%%%%% START OF SUPPLEMENT %%%%%%%%%%%%%%%

% Figures, tables, equations and pages in the supplement are numbered S1, S2 etc.
\renewcommand{\thefigure}{S\arabic{figure}}
\renewcommand{\thetable}{S\arabic{table}}
\renewcommand{\theequation}{S\arabic{equation}}
\renewcommand{\thepage}{S\arabic{page}}
\setcounter{figure}{0}
\setcounter{table}{0}
\setcounter{equation}{0}
\setcounter{page}{1} % not 0 as \newpage already started a supplementary page
% References continue the numbering from the main text.

%%%%%%%%%%%%%%%% SUPPLEMENT TITLE PAGE %%%%%%%%%%%%%%%
\clearpage
\onecolumngrid
\begin{center}
\section*{Supplemental Information}
\maketitle

\end{center}

\noindent {This PDF file includes:}

\noindent Supplementary Text\\
Figures S1 to S10\\

%%%%%%%%%%%%%%%% SUPPLEMENTARY TEXT %%%%%%%%%%%%%%%
\clearpage
\twocolumngrid

\section*{Supplementary Text}

\subsection*{Comprehensive symmetry analysis} 

To investigate the origin of the photocurrents, we model the time-dependent dynamics with both analytic and \textit{ab-initio} methods.
In figure~\ref{Fig3_total} we present the CNR analysis discussed around figure~\ref{Fig3} again, highlighting the local symmetries, for clarity.
Figures~\ref{Fig3_total}a--l show the residual conduction band distributions for the COR, CNR, and TD waveforms obtained from TDDFT. 
Analyzing all six K and K' points allows us to look at the dynamical selection rules for the respective waveforms globally, while the zoom-ins (figs.~\ref{Fig3_total}g--l) highlight the intra-valley symmetries.

We analyze the dynamical selection rules by the residual conduction band populations for two orthogonal phases of both the COR, CNR, and TD waveforms.
After excitation with the COR waveform (figs.~\ref{Fig3_total}a--b) we can see one mirror axis present in each of the graphene band structures corresponding to the waveform symmetry.
For \pr = $\pi/2$ (fig.~\ref{Fig3_total}a), we observe a left/right symmetry in the conduction band population, resulting in no current generation, with a dominate symmetry axis through M--$\Gamma$--M', addressing the K and K' points equivalently, rendering  $\rho^\mathrm{c}$ K and K’ symmetric.
The zoom-in (fig.~\ref{Fig3_total}g) highlights the crescent features showing this vertical mirror axis between the K and K' points.
When \pr = $0$ (figs.~\ref{Fig3_total}b, h), our waveform's principle symmetry axis changes to the K--$\Gamma$--K' axis. 
This disconnects the K and K' points and allows intra-valley momentum imbalances and a net current. 

For the excitation with a CNR waveform (figs.~\ref{Fig3_total}c , d), similar arguments can be made, now with the waveforms exhibiting two additional minor mirror axes (dashed orange lines) corresponding to the three-fold rotational symmetry of the waveform.
Importantly, the time reversal symmetry (purple color bar) is coupled to the primary mirror axis (green).
Critically, shown in the \pr = $\pi/2$ waveform, the symmetry axis connects K and K' (a M-$\Gamma$-M' axis), resulting in the inability to generate a valley polarization.
With \pr = $0$, the waveform's symmetry axis (K-$\Gamma$-K') now disconnects the K and K' points, resulting in possible valley polarization. 
The zoom-ins of the CNR population (fig.~\ref{Fig3_total}i, j) reveal the local symmetries (dashed orange lines) within the K and K' valleys.

After excitation with the TD waveform (figs.~\ref{Fig3_total}e, f) we can see one mirror axis present in each of the graphene band structures corresponding to the waveform symmetry.
For \pr = $\pi/2$ (fig.~\ref{Fig3_total}e), we observe a left/right symmetry in the conduction band population, resulting in no current generation, with a dominate symmetry axis through M--$\Gamma$--M', addressing the K and K' points equivalently, rendering  $\rho^\mathrm{c}$ K and K’ symmetric.
The zoom in (fig.~\ref{Fig3_total}k) highlights the crescent features showing this vertical mirror axis between the K and K' points.
When \pr = $0$ (figs.~\ref{Fig3_total}f, l), our waveform's principle symmetry axis changes to the K--$\Gamma$--K' axis. 
This disconnects the K and K' points and allows intra-valley momentum imbalances and a net current. 

To understand the emergence of these characteristic patterns (seen in figs.~\ref{Fig3_total}h, g, and l) and the underlying valley control on a sub-cycle time scale, it is insightful to discuss a picture of time-dependent inter- and intraband electron dynamics in the undressed band structure. 
We highlight these dynamics for three exemplary electrons with initial momenta equidistant to the respective K and K' point (figs.~\ref{Fig3_total}h, j, l, colored circles). 
Figures~\ref{Fig3_total}n, p, r, show the respective sub-cycle evolution of the interband dipole coupling $V^\mathrm{cv}(t)$ induced by the COR, CNR, and TD waveform, respectively (at \pr$=0$). 
The stark inter-valley differences in $|V^\mathrm{cv}|$ occurring during the center optical cycle (figs.~\ref{Fig3_total}m, o, q; shaded areas) become clear with a fundamental field strength of ${1}$~V/nm applied here for visibility.

The COR waveform drives the right marked electron in the K' valley (fig.~\ref{Fig3_total}h) close to the Dirac point, resulting in a large excitation probability (fig.~\ref{Fig3_total}n), whereas it is suppressed for the other two marked electrons, which are driven away from the Dirac point. Hence all three exemplary electrons are shown to assist in the generation of intra-valley momentum asymmetry, observed in the crescent shaped distribution of fig.~\ref{Fig3_total}h and measured as a large current (Fig.~\ref{Fig2}b). The three marked electrons in the K valley contribute likewise.

In contrast, the CNR waveform drives all three exemplary electrons in the K' valley (fig.~\ref{Fig3_total}j) equally close to the Dirac point. 
At the K point, the same three electrons come close to the Dirac point only for a highly confined duration at the trefoil peaks, lowering their excitation probability. 
In the dipole coupling (fig.~\ref{Fig3_total}p) this is reflected in a larger amplitude of $|V^\mathrm{cv}|$ for each electron around K' compared to K. 
Importantly, these valley-specific differences in interband coupling result in a valley-polarized conduction band population. 
In addition, since the relative amplitudes of the marked points are not exchanged between K and K' (as is the case for the COR waveform, fig.~\ref{Fig3_total}n), there is no intra-valley momentum asymmetry. 
This allows us to identify the drastic reduction of the measured current (Fig.~\ref{Fig2}b) as an experimental generation of valley-polarized graphene.

The TD waveform drives the right marked electron in the K' valley (fig.~\ref{Fig3_total}l) equivalently close to the Dirac point, resulting in an approximately equal excitation probability (fig.~\ref{Fig3_total}r).
The three exemplary electrons in the K' valley behave similarly to the CNR case. 
Electrons around the K valley are driven differently. The electron on the mirror plane is driven close to the dirac point, while the other two are not driven to the dirac point (fig.~\ref{Fig3_total}r).
All three exemplary electrons in the K valley are shown to assist in the generation of intra-valley momentum asymmetry, observed in the crescent shaped distribution of fig.~\ref{Fig3_total}l and measured as a large current (Fig.~\ref{Fig2}b).
The intravalley-asymmetry present in only one of the two valleys produces valley current.

\subsection*{Emergent Chern numbers}
To verify that the two pulses can be split into a constant dressing pulse and a drive pulse, we calculate the emerging Chern numbers for all three combinations of driving fields using the Hamiltonian from the high frequency expansion~\cite{Eckardt2015}. The Berry curvature is calculated identically to the Methods section~\cite{Xiao2010}. We present the three cases in figure~\ref{HFE} and highlight that the topological state is equivalent for all three cases. 

Further, for the specific case of the anomalous Hall current, we compare the Floquet band structure of only the dressing ($\omega$) pulse with the dressing and drive ($\omega+2\omega$) pulses combined (fig.~\ref{BerryComp}). We see that the Berry curvature is largely determined by the dressing ($\omega$) pulse while the driving ($2\omega$) pulse has less impact on the Berry curvature due to the weaker vector potential and higher frequency. This is in line with the results from the high frequency expansion shown above.

\subsection*{ARPES Simulation}
We calculate the time and angle resolved photoelectron spectrum (Tr-ARPES) of graphene (fig.~\ref{ARPES}) under identical light dressing conditions to the experiment, with conduction band doping. The results are calculated ab-initio with state-of-the-art simulations, and including doping and electron-electron interactions within TDDFT (with the same approach employed in ref.~\cite{microarpes_2024}). We see gap sizes and Floquet bands that match those predicted by the model Hamiltonain discussed earlier, clearly demonstrating in this regime, with near-infrared driving presented in this paper, an FTI state is created with a topological gap that should be measurable.

\subsection*{Field strength photocurrent scaling}
We measure the two-color photocurrent as a function of combined field strength for COR, CNR, and linear (parallel) polarizations. We plot the scaling in figure~\ref{Scaling} in both linear (a) and log (b) scales.

\subsection*{Pulse Characterization}
We characterize the pulses after the BiBO crystal by second harmonic generation frequency resolved optical gating (SHG-FROG). We obtain a 213~fs pulse for the fundamental and a 110~fs pulse for the second harmonic, as shown in figure~\ref{FROG}.

%%%%%%%%%%%%%%%% SUPPLEMENTARY FIGURES %%%%%%%%%%%%%%%

\begin{figure*}[h!]
\includegraphics[width=\linewidth]{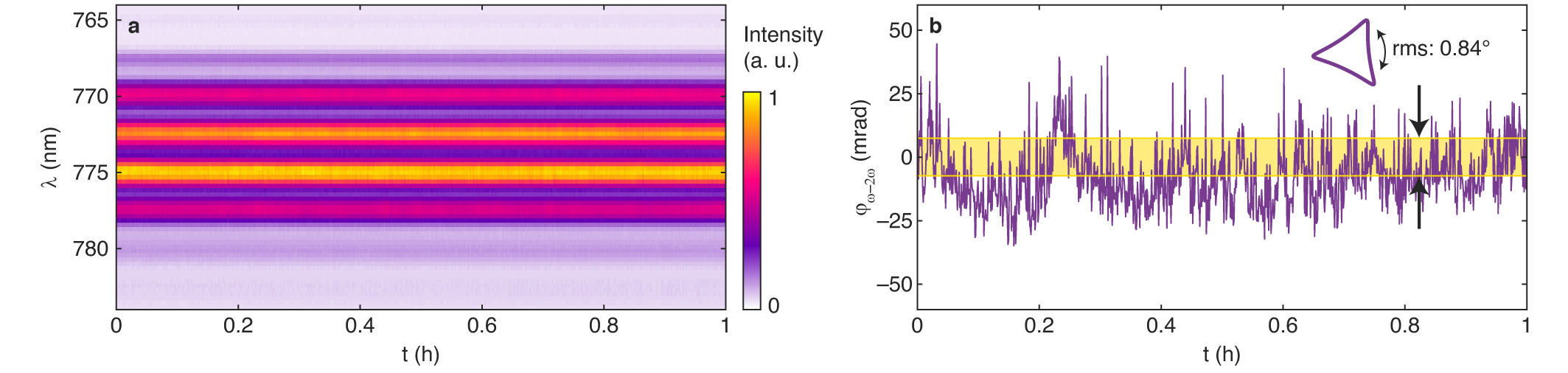}
    \caption{\textbf{Phase stability of the collinear interferometer.} \textbf{a}, Interference fringes at $2\omega$ measured every {35}~{ms} for {1}~{hour} with a spectrometer integration time of 10~{ms}. \textbf{b}, Reconstructed jitter of \pr\ obtained from a cosine square fit, yielding a root mean square of {14.6}~{mrad} ({0.84}{$^\circ$}) for the angular stability of the bichromatic Lissajous figures (yellow shading).}
    \label{FigS1}
\end{figure*}

\begin{figure*}[h!]
\includegraphics[width=\linewidth]{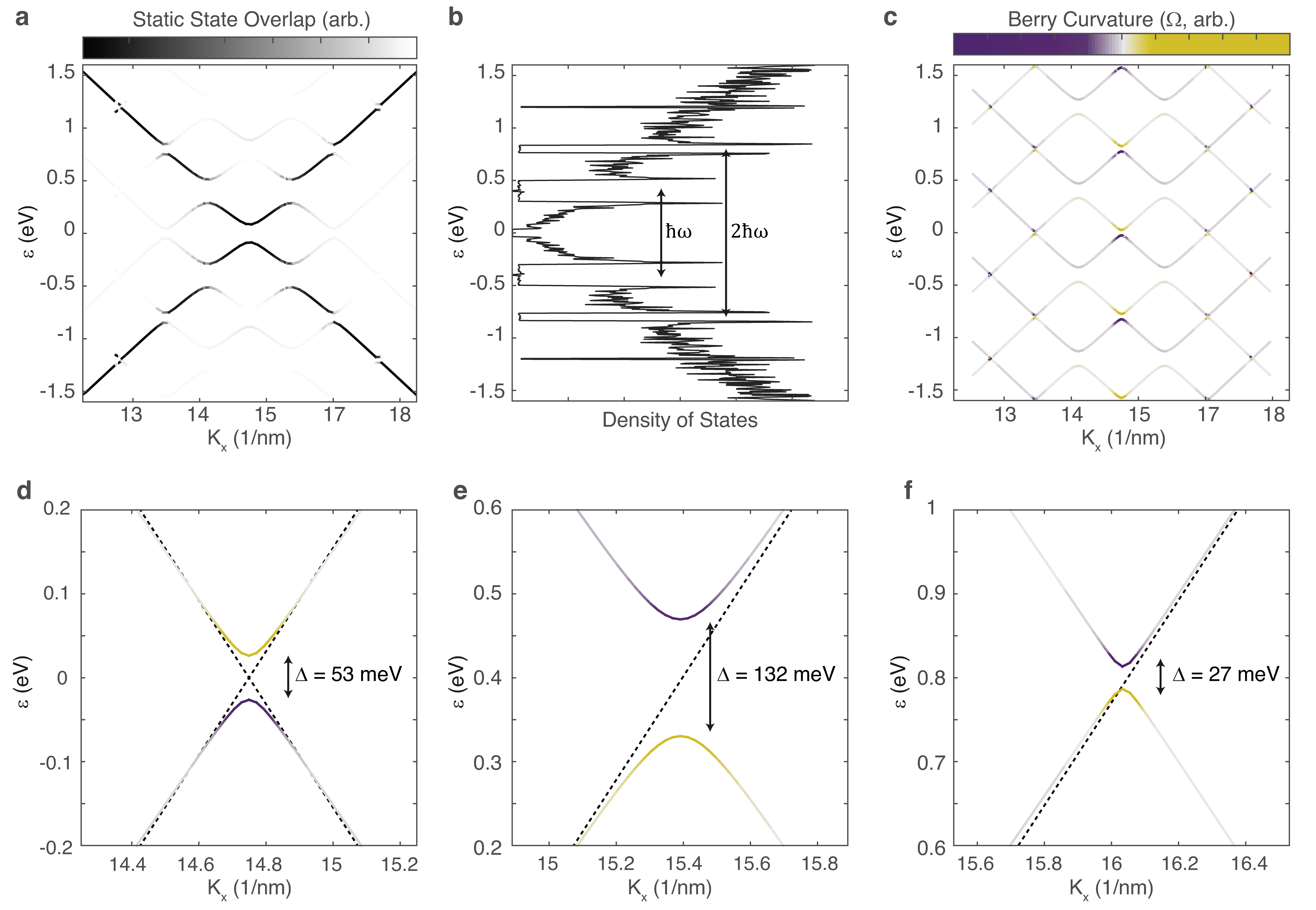}
    \caption{{\color{black}\textbf{Floquet topological insulating state in graphene.} \textbf{a} Calculated Floquet band structure for graphene under experimental illumination conditions, static state overlap calculated from \cite{Oka2009,Sato2019}. \textbf{b} Density of states for the dressed band structure in (a) showing bandgaps at the Dirac point, one, and two-photon resonances. \textbf{c} Berry curvature of the dressed band structure\cite{Xiao2010}. \textbf{d--f} Zoom in of the topological band gaps and Berry curvature at the K point (\textbf{d}), one photon resonance (\textbf{e}, 10x Berry curvature),  and two photon resonance (\textbf{f}, 10x Berry curvature).}}
    \label{FigFloquet}
\end{figure*}

\begin{figure*}[h!]
\includegraphics[width=.7\linewidth]{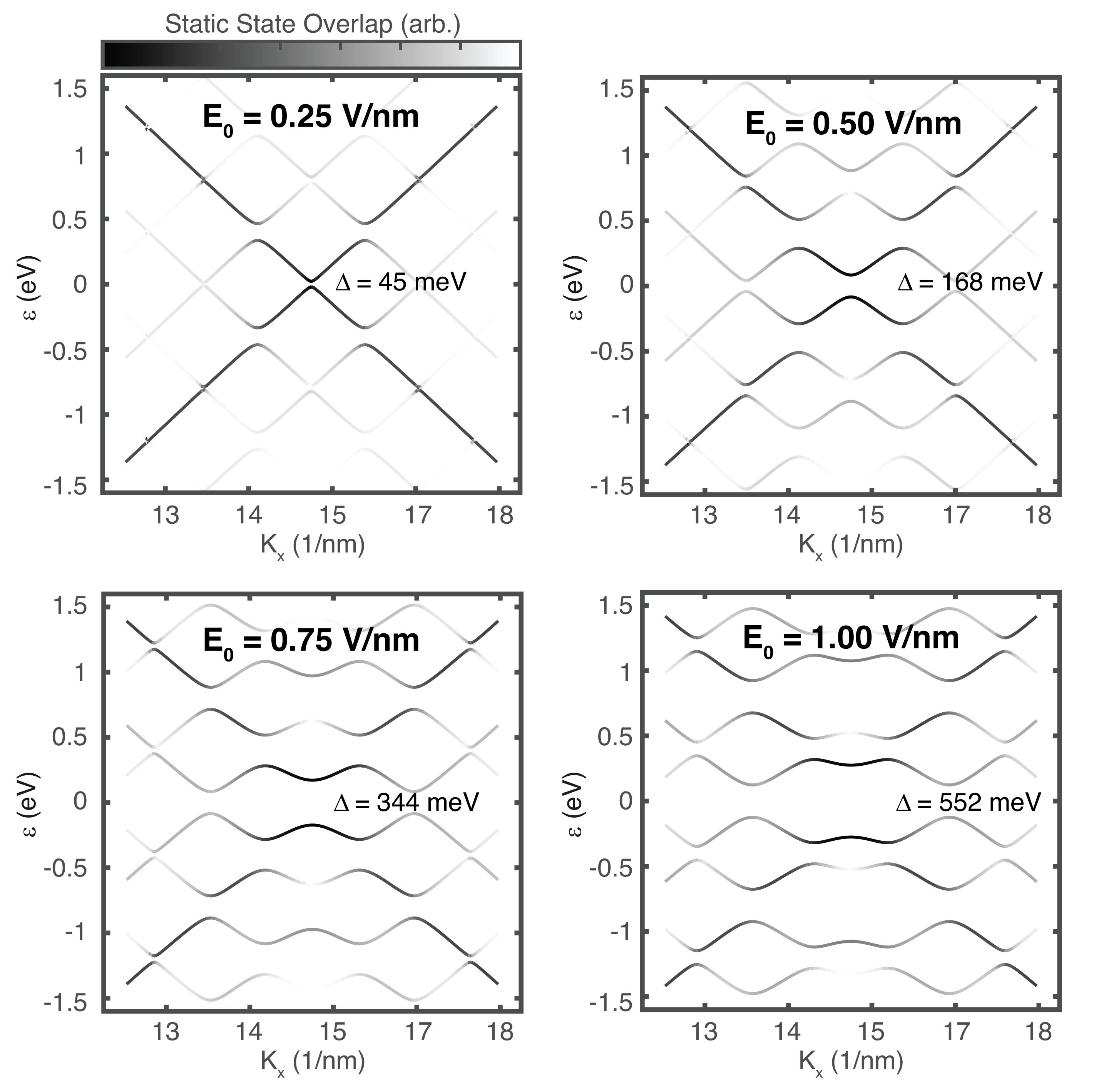}
    \caption{{\color{black}\textbf{Large bandgap Floquet topological insulating states in graphene.} Calculated Floquet band structure for graphene under a variety of experimentally feasible field strengths at 1550~nm~\cite{Lesko2021,Lesko2022}, static state overlap calculated from \cite{Oka2009,Sato2019}.}}
    \label{FigFloquetScaling}
\end{figure*}

\begin{figure*}[h!]
\includegraphics[width=0.5\linewidth]{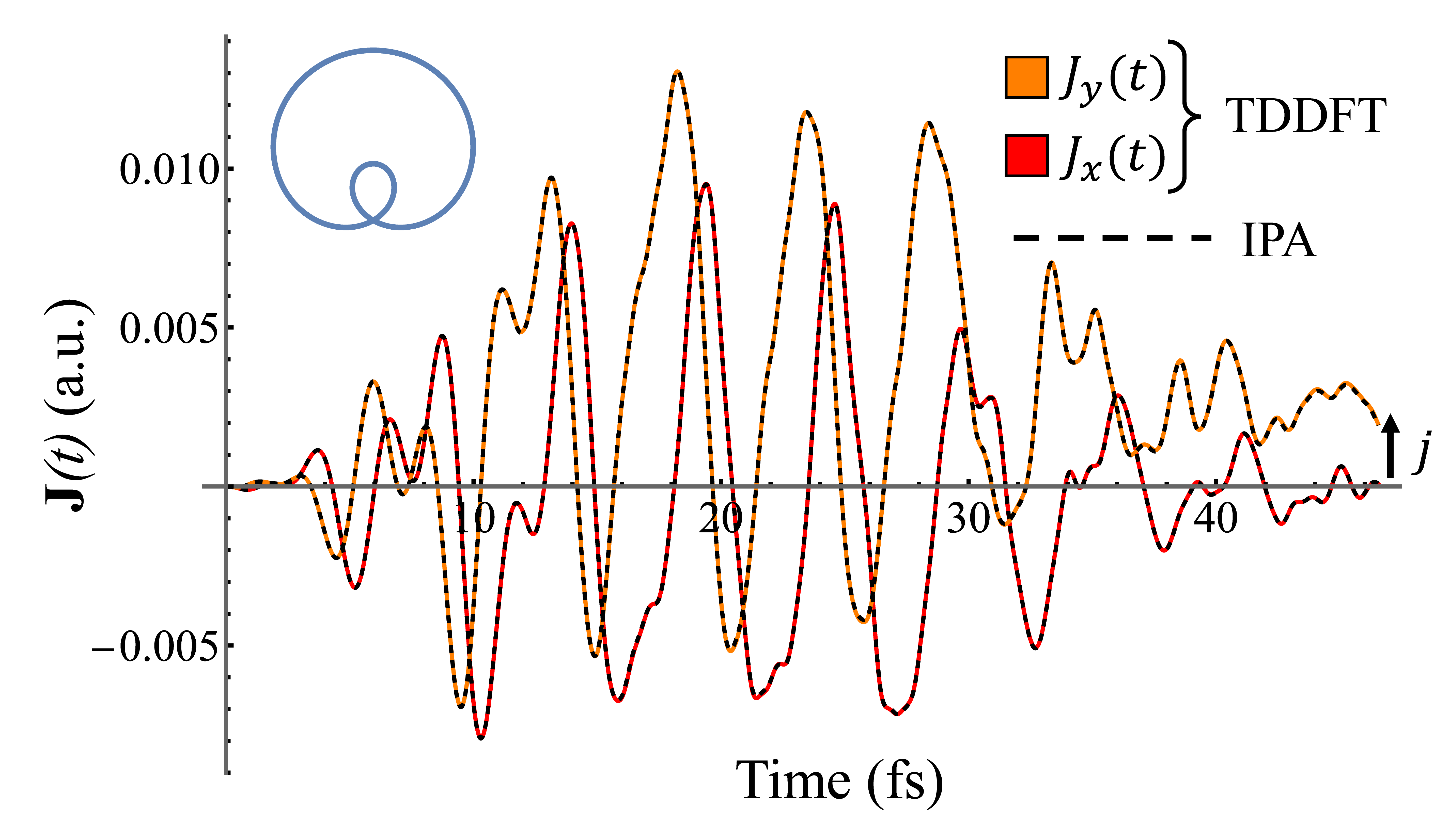}
    \caption{\textbf{Comparison of calculated current expectation value}, $\textbf{J}(t)$ from eq.~\ref{eq:proj}, with a full TDDFT calculation, or within the IPA with frozen electronic interactions. The system's response is almost identical in both cases, indicating the interactions are negligible. Calculated for co-circular case at a high laser power of $10^{11}$~W/cm$^{2}$ with 1550~nm driving.}
    \label{FigS2}
\end{figure*}

\begin{figure*}[h!]
\includegraphics[width=0.7\linewidth]{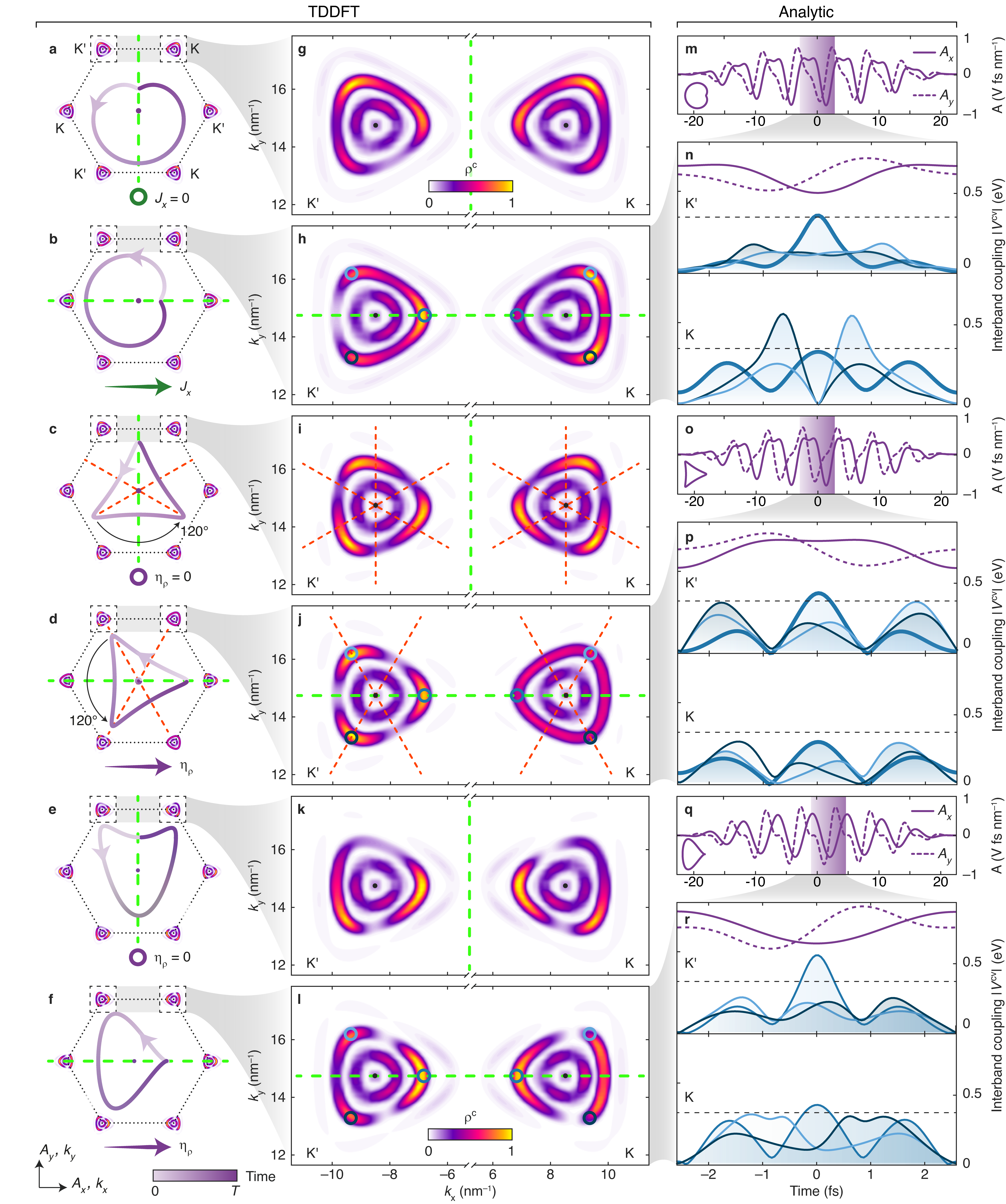}
    \caption{\textbf{Sub-cycle dynamical symmetry selection rules.} \textbf{a--d}, Temporal shape and symmetries of the vector potential Lissajous figures (purple lines) and the resulting symmetries of residual momentum distributions $\rho^\mathrm{c}$ imprinted to the first Brillouin zone (dotted hexagons) for the COR (\textbf{a, b}), CNR (\textbf{c, d}), and  TD (\textbf{e, f}) waveform for $\varphi_\mathrm{\omega-2\omega}=\pi/2$ and $0$ respectively. The momentum distributions are computed with TDDFT (see Methods for details). Green (orange) dashed lines indicate global (valley-specific) symmetry axes, determined by both waveform and time periodicity (see purple color bar). \textbf{g, l}, Close-ups of $\rho^\mathrm{c}$ around K and K' as indicated in \textbf{b}, \textbf{d}, and \textbf{f}. 
    For all cases, an eight-cycle pulse (super-sine envelope, see vector potential in \textbf{m}, \textbf{o}, and \textbf{q}) is applied with {1550}~{nm} fundamental wavelength with a peak field strength $E_{0}={0.27}$~{V/nm} and 75~\% $2\omega$ field admixture. 
    \textbf{m}, \textbf{o}, \textbf{q}, COR (\textbf{m}), CNR (\textbf{o}), and TD (\textbf{q}) vector potential waveform for $\varphi_{\omega-2\omega}=0$. 
    The shaded areas indicate the center optical cycle respectively. \textbf{n}, \textbf{p}, \textbf{r}, Temporal profile of the interband dipole coupling $|V^\mathrm{cv}|$~\cite{Weitz2024} induced by the waveforms shown in \textbf{m}, \textbf{o}, and \textbf{q}, respectively. 
    The different shades of blue refer to initial momenta in the K and K' valley as marked with circles in panels \textbf{h}--\textbf{l}. The dashed black guide lines compare the $|V^\mathrm{cv}|$ amplitudes at K and K'.}
    \label{Fig3_total}
\end{figure*}

\begin{figure*}[h!]
    \includegraphics[width=\linewidth]{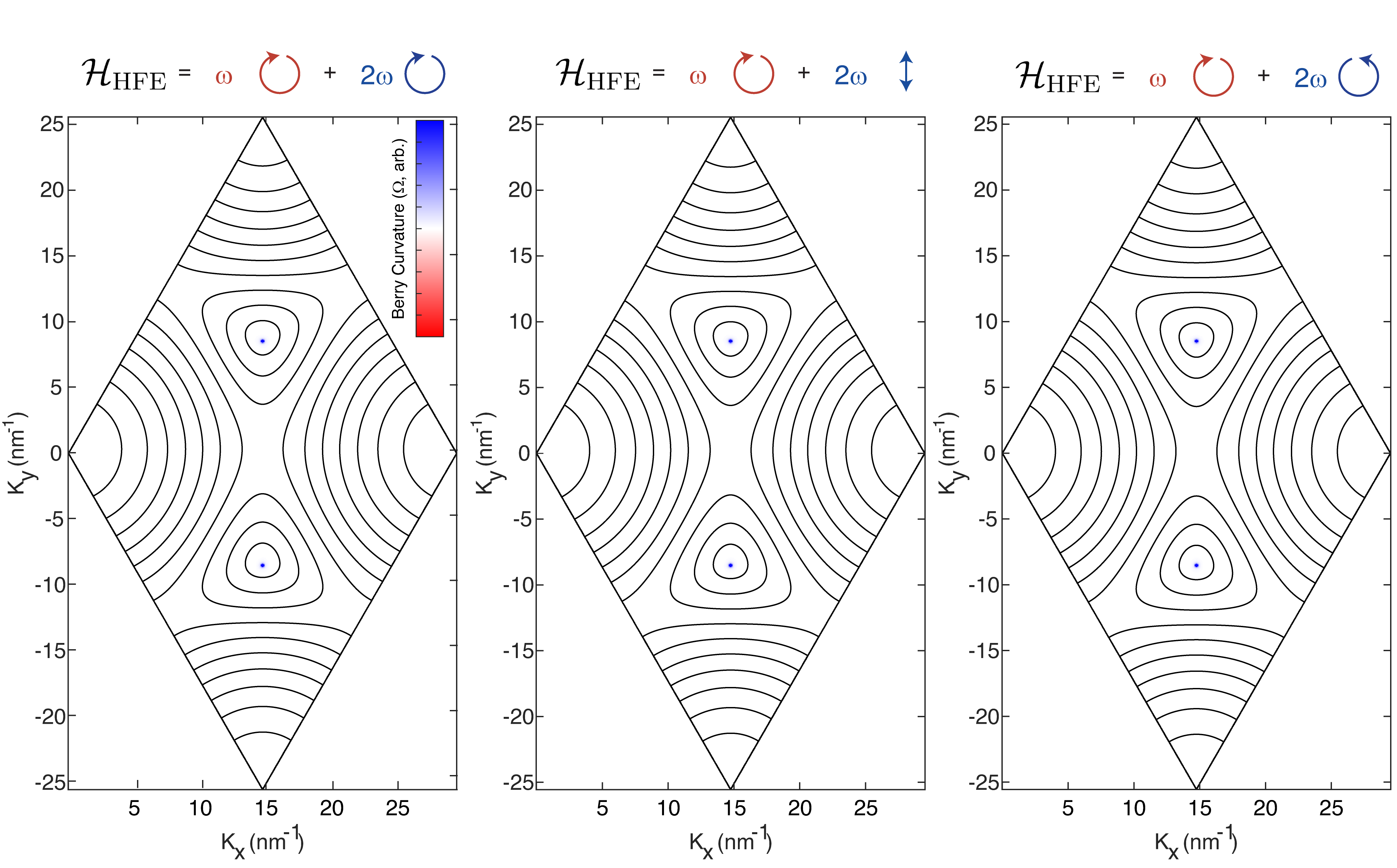}
    \caption{{\color{black}\textbf{Emerging Chern numbers from two-color driving.} We plot the Berry curvature of the full Brillouin zone for each of the two color driving cases using the high frequency expansion. The Chern number is equivalent for all three of the cases.}}
    \label{HFE}
\end{figure*}

\begin{figure*}[h!]
\includegraphics[width=.5\linewidth]{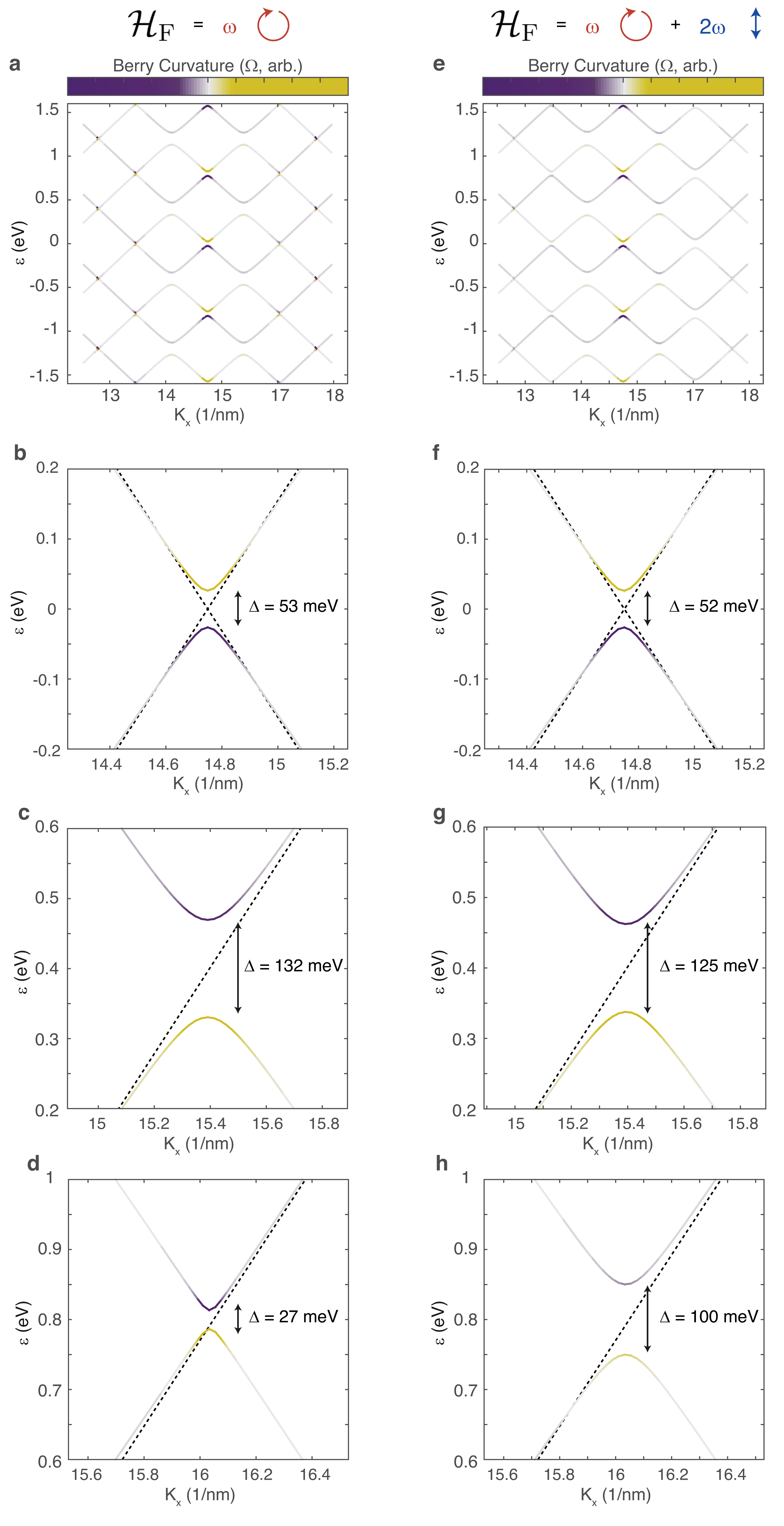}
    \caption{{\color{black}\textbf{Comparison of FTI states.} We plot the Floquet band structure and Berry curvature for $\omega$ only (\textbf{a--d}) as well as $\omega + 2\omega$ (\textbf{e--h}).}}
    \label{BerryComp}
\end{figure*}

\begin{figure*}[h!]
\includegraphics[width=.5\linewidth]{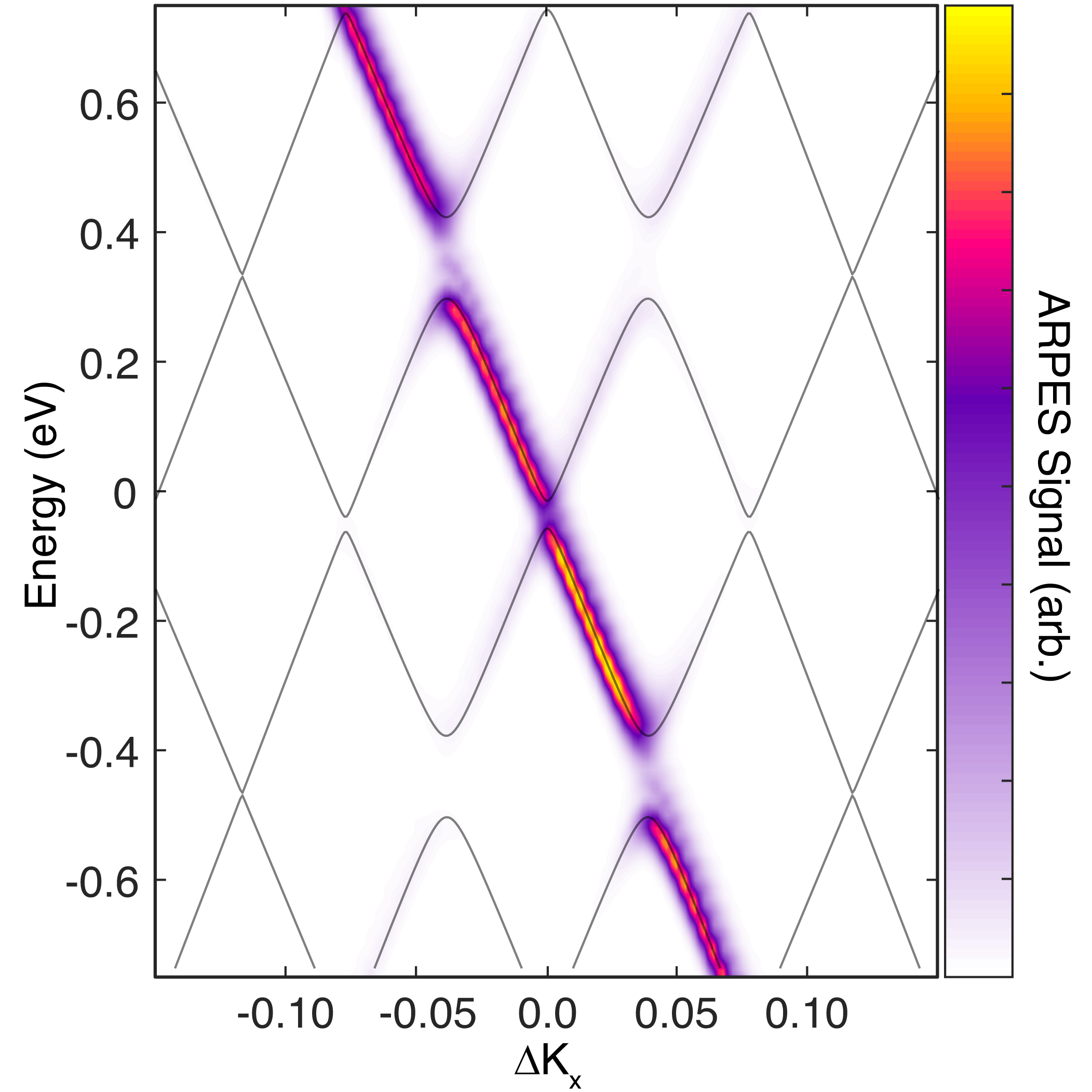}
    \caption{{\color{black}\textbf{Calculated ARPES spectra of dressed graphene.} ARPES spectra of n-doped graphene dressed with an optical field identical to the one in this manuscript showing an evident bandgap at K and the first avoided crossing.}}
    \label{ARPES}
\end{figure*}

\begin{figure*}[h!]
\includegraphics[width=.5\linewidth]{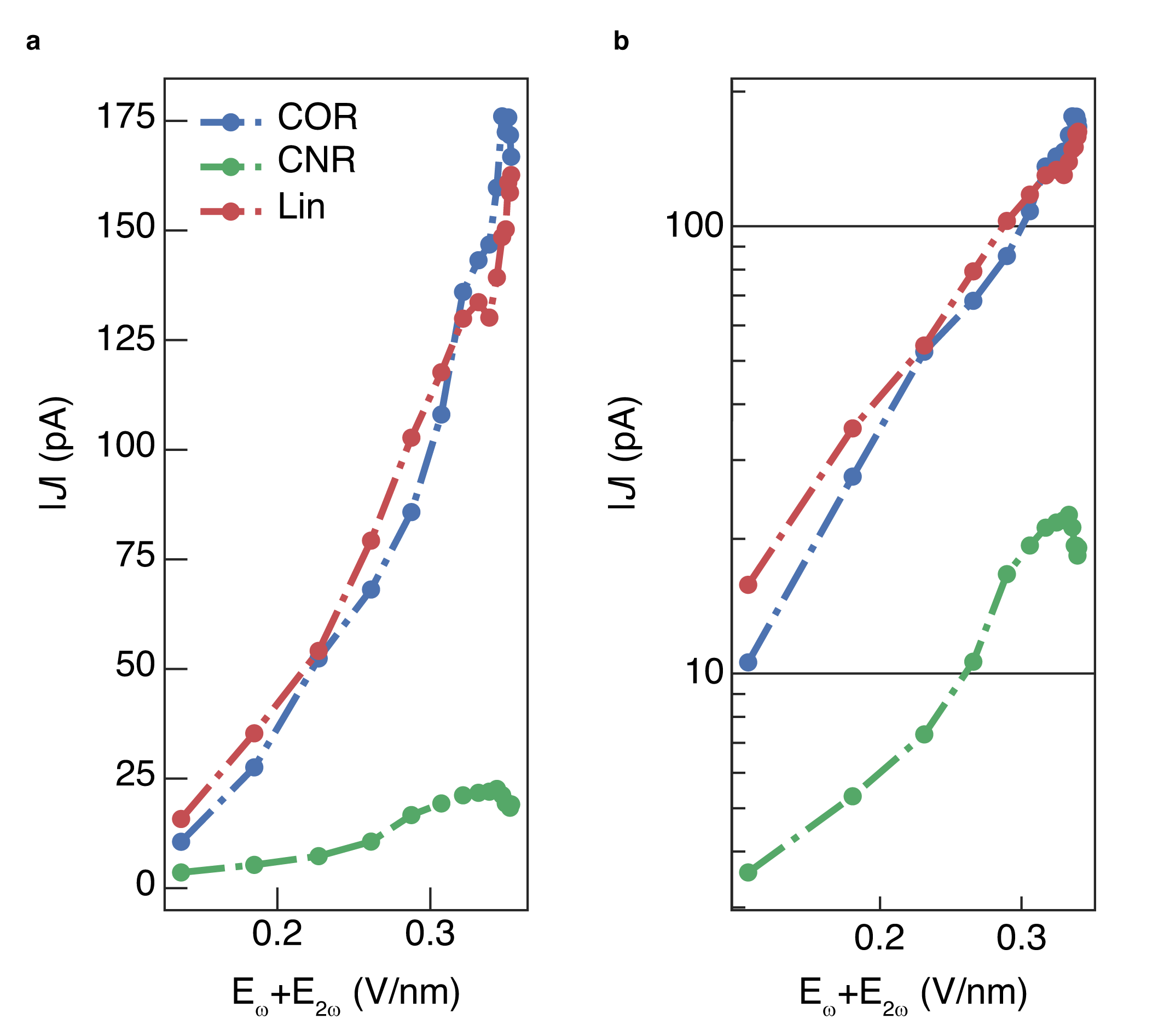}
    \caption{{\color{black}\textbf{Field strength scaling.} The two-color current scaling as a function of the $\omega + 2\omega$ field strength on both linear (\textbf{a}) and log (\textbf{b}) scale.}}
    \label{Scaling}
\end{figure*}

\begin{figure*}[h!]
\includegraphics[width=\linewidth]{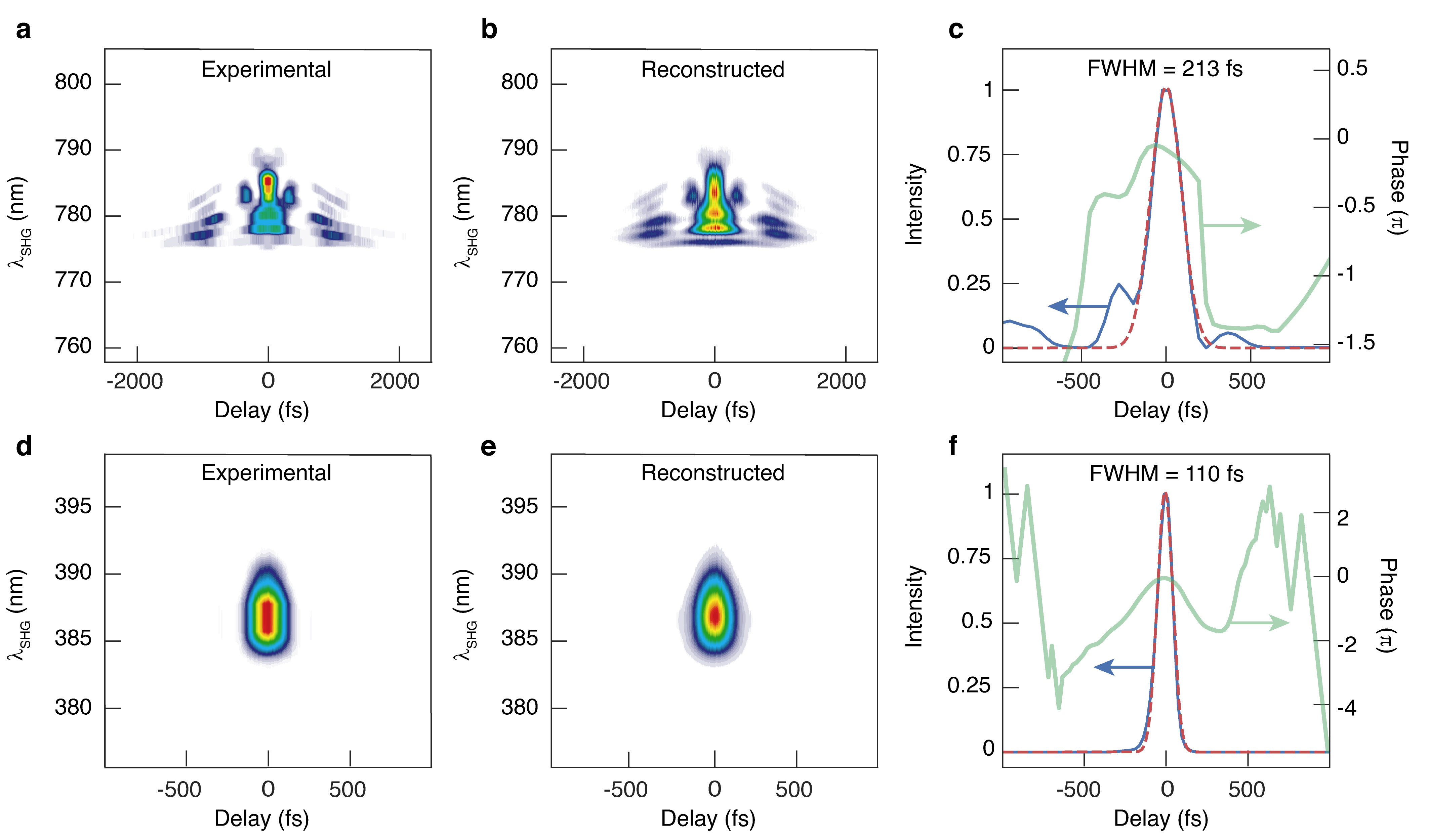}
    \caption{{\color{black}\textbf{Optical Pulse Characterization} We measure the SHG-FROG spectrogram for $\omega$ (\textbf{a}) and reconstruct (\textbf{b}) a 213~fs pulse ((\textbf{c})). The same is done for the $2\omega$ pulse (\textbf{d--f}) revealing a 110~fs pulse.}}
    \label{FROG}
\end{figure*}

\end{document}